\documentclass[aop,citesort,MSNbibl,dvips]{arximspdf}

\doi{10.1214/10-AOP544}
\volume{38}
\issue{6}
\pubyear{2010}
\firstpage{2224}
\lastpage{2257}

\makeatletter

\newtheorem{theo}{Theorem}
\newproclaim{defini}{Definition}
\newtheorem{proposi}{Proposition}
\newtheorem{lemma}{Lemma}
\newtheorem{coro}{Corollary}
\newproclaim{Remark}{Remark}

\newcommand{\CC}{{\mathbb C}}
\newcommand{\NN}{{\mathbb N}}
\newcommand{\RR}{{\mathbb R}}
\newcommand{\TM}{{\mathbb T}}
\newcommand{\ZZ}{{\mathbb Z}}
\newcommand{\IM}{{\mathbb I}}

\newcommand{\Dd}{\mathcal{D}}
\newcommand{\PP}{\mathbf{P}}
\newcommand{\EE}{\mathbf{E}}
\newcommand{\Ff}{\mathcal{F}}
\newcommand{\Gg}{\mathcal{G}}
\newcommand{\Ww}{\mathcal{W}}
\newcommand{\Ss}{\mathcal{S}}
\newcommand{\Oo}{\mathcal{O}}
\newcommand{\Tr}{\operatorname{Tr}}
\newcommand{\Tt}{\mathcal{T}}
\newcommand{\Mm}{\mathcal{M}}
\newcommand{\Jj}{\mathcal{J}}
\newcommand{\Ll}{\mathcal{L}}
\newcommand{\Qq}{\mathcal{Q}}
\newcommand{\Uu}{\mathcal{U}}
\newcommand{\Vv}{\mathcal{V}}
\newcommand{\dis}{{\mathrm{dis}}}
\newcommand{\one}{\mathbf{1}}
\newcommand{\nul}{\mathbf{0}}

\newcommand{\diag}{\operatorname{diag}}

\newcommand{\ppp}{\mathfrak{p}}
\newcommand{\ggm}{\mathfrak{g}}
\newcommand{\bbb}{\mathfrak{b}}
\newcommand{\uuu}{\mathfrak{u}}
\newcommand{\vvv}{\mathfrak{v}}
\newcommand{\rrr}{\mathfrak{r}}

\newcommand{\Kkk}{\mathfrak{K}}
\newcommand{\Hhh}{\mathfrak{H}}
\newcommand{\Ttt}{\mathfrak{T}}

\makeatother

\begin{document}
\begin{frontmatter}

\title{Random Lie group actions on compact manifolds: A~perturbative
analysis\thanksref{T1}}
\runtitle{Random Lie group actions}

\thankstext{T1}{Supported by the DFG.}

\begin{aug}
\author[A]{\fnms{Christian} \snm{Sadel}} and
\author[A]{\fnms{Hermann} \snm{Schulz-Baldes}\corref{}\ead[label=e1]{schuba@mi.uni-erlangen.de}}
\runauthor{C. Sadel and H. Schulz-Baldes}
\affiliation{Universit\"at Erlangen--N\"urnberg}
\address[A]{Department Mathematik\\
Universit\"at Erlangen--N\"urnberg\\
Bismakrstrasse 1 1/2\\
Erlangen\\
Germany\\
\printead{e1}} 
\end{aug}

\received{\smonth{2} \syear{2008}}
\revised{\smonth{8} \syear{2009}}

%
\begin{abstract}
A random Lie group action on a compact manifold generates a discrete
time Markov process. The main object of this paper is the evaluation of
associated Birkhoff sums in a regime of weak, but sufficiently
effective coupling of the randomness. This effectiveness is expressed
in terms of random Lie algebra elements and replaces the transience or
Furstenberg's irreducibility hypothesis in related problems. The
Birkhoff sum of any given smooth function then turns out to be equal to
its integral w.r.t. a unique smooth measure on the manifold up to
errors of the order of the coupling constant. Applications to the
theory of products of random matrices and a model of a disordered
quantum wire are presented.
\end{abstract}

%

\end{frontmatter}

\section{Main results, discussion and applications}\label{sec1}

This work provides a perturbative calculation of invariant measures
for a class of Markov chains on continuous state spaces and shows that
these perturbative measures are unique and smooth.
Let us state the main result right away in detail, and then
place it into context with other work towards the end of this section
and explain our motivation to
study this problem.

Suppose given a Lie group $\Gg\subset\operatorname{GL}(L,\CC)$,
a compact, connected, smooth Riemannian
manifold $\Mm$ without boundary and a smooth, transitive group action
$\cdot\dvtx \Gg\times\Mm\to\Mm$. Thus, $\Mm$ is a homogeneous space.
Furthermore, let $\Tt_{\lambda,\sigma}\in\Gg$ be a family of group
elements depending on
a coupling constant $\lambda\geq0$ and a parameter
$\sigma$ varying in some probability space $(\Sigma,\mathbf{p})$, which is
of the following form:
%
%
\begin{equation}
\label{eq-T}
\Tt_{\lambda,\sigma}
=
\mathcal{R}
\exp\Biggl( \sum_{n=1}^\infty\lambda^n \mathcal{P}_{n,\sigma
} \Biggr) ,
\end{equation}
where $\mathcal{R}\in\Gg$ and $\mathcal{P}_{n,\sigma}$ are measurable
maps on $\Sigma$ with
compact image in the Lie algebra $\mathfrak{g}$
of $\Gg$ such that
%
%
\begin{equation}
\label{eq-taylorradius}
\limsup_{n\to\infty} \sup_{\sigma\in\Sigma}
(\|\mathcal{P}_{n,\sigma}\|)^{{1/n}}
< \infty
\end{equation}
for some norm on $\ggm$. This implies that $\Tt_{\lambda,\sigma}$ is
well defined and analytic in $\lambda$ for
$\lambda$ sufficiently small.
The expectation value of the first-order term $\mathcal{P}_{1,\sigma}$
will be denoted by $\mathcal{P}=\int\mathbf{p}(d\sigma) \mathcal
{P}_{1,\sigma}$.

Let us consider the product probability space
$(\Omega,\PP)=(\Sigma^\NN,\mathbf{p}^\NN)$.
Associated to $\omega=(\sigma_n)_{n\in\NN}\in\Omega$, there is
a sequence $(\Tt_{\lambda,\sigma_n})_{n\in\NN}$ of group elements.
An $\Mm$-valued Markov process $x_n(\lambda,\omega)$
with starting point $x_0 \in\Mm$ is defined iteratively by
%
%
\begin{equation}
\label{eq-Xn}
x_n(\lambda,\omega)
=
\Tt_{\lambda,\sigma_n} \cdot x_{n-1}(\lambda,\omega) .
\end{equation}
The averaged Birkhoff sum of a complex function $f$ on $\Mm$ is
%
%
\begin{equation}
\label{eq-avBirk}
I_{\lambda,N}(f)
=
\EE_\omega\frac{1}{N} \sum_{n=0}^{N-1} f(x_n(\lambda,\omega))
=
\frac{1}{N} \sum_{n=0}^{N-1} (T_\lambda^n f)(x_0)
,
\end{equation}
where in the second expression we used the Markov transition operator
$(T_\lambda f)(x)=\EE_\sigma(f(\Tt_{\lambda,\sigma}\cdot x))$.
Here and below, expectation values w.r.t. $\PP$ (or $\mathbf{p}$)
will be denoted by $\EE$ (or $\EE_\omega$ and $\EE_\sigma$).
Next, recall that an invariant measure
$\nu_\lambda$ on $\Mm$ is defined by the property
$\int\nu_\lambda(dx) f(x)=\int\nu_\lambda(dx) (T_\lambda f)(x)$.
The operator ergodic theorem \cite{Kal}, Theorem 19.2, then states that
$I_{\lambda,N}(f)$ converges almost surely (in $x_0$)
w.r.t. any invariant measure $\nu_\lambda$ and for
any integrable function $f$. In the case that $\Mm$ is a projective
space and the action is matrix multiplication, one is in the world of
products of random matrices. If then the group generated
by $\Tt_{\lambda,\sigma}$, with $\sigma$ varying in the support of
$\mathbf{p}$, is noncompact and strongly irreducible, Furstenberg, Guivarch
and Raugi have
proved \cite{Fur,GR,BL} that there is a unique invariant measure
$\nu_\lambda$ which is, moreover, H\"older continuous \cite{BL}. To
our best knowledge, little seems to be known in more general
situations and also concerning the absolute continuity of $\nu_\lambda$
(except if $\mathbf{p}$ is absolutely continuous \cite{Lia}, for some
and under supplementary hypothesis \cite{ST,CK}).

Let $\mathbf{p}_1$ be the distribution of
the random variable $\mathcal{P}_{1,\sigma}$ on the Lie algebra $\ggm
$, that is,
for any measurable $\bbb\subset\ggm$ one has
$\mathbf{p}_1(\bbb)=\mathbf{p}(\{\mathcal{P}_{1,\sigma}\in\bbb\})$.
We are interested in a perturbative calculation of $I_{\lambda,N}(f)$
in $\lambda$ for smooth functions $f$ with rigorous control
on the error terms. This can be achieved if the
support of $\mathbf{p}_1$ is large enough in the following sense.
First, let us focus on the special case $\mathcal{R}=\one$ and
$\mathcal{P}=0$.
%
\begin{theo}
\label{theo-main2}
Let $\Tt_{\lambda,\sigma}$ be of the form (\ref{eq-T}) and
assume $\mathcal{R}=\one, \mathcal{P}=\EE(\mathcal{P}_{1,\sigma
})=0$. Let
$x_n$ be the associated Markov process on $\Mm$ as given
by (\ref{eq-Xn})
and let $\vvv=\operatorname{Lie} (\operatorname{supp}(\mathbf{p}_1) )$
be the smallest Lie subalgebra of $\ggm$ that
contains the support of $\mathbf{p}_1$.
Recall that $\mu(dx)$ denotes the Riemannian volume measure on
$\Mm$.

\textup{Coupling hypothesis}:
Suppose that the smallest subgroup $\Vv$ of $\Gg$ containing
$\{\exp(\lambda\mathcal{P}), \mathcal{P}\in\vvv, \lambda\in
[0,1]\}$
acts transitively on $\Mm$. (This is a Lie subgroup with Lie
algebra $\vvv$, but it may not be a submanifold.)

Then there is a sequence of smooth functions $\rho_m$ with
$\int_\Mm d\mu\rho_m=\delta_{m,0}$ and $\rho_0 > 0$
$\mu$-almost surely, such that for any $M\in\NN$ and
any function $f\in C^\infty(\Mm)$, one obtains
%
%
\begin{equation}
\label{eq-Birkhoffresult2}
I_{\lambda,N}(f)
=
\sum_{m=0}^M \lambda^m \int_\Mm\mu(dx) \rho_m(x) f(x)
+
\Oo\biggl( \frac{1}{N\lambda^2},\lambda^{M+1} \biggr) .
\end{equation}
\end{theo}
%

Here, the\vspace*{1pt} expression $\Oo( \frac{1}{N\lambda^2},\lambda^{M+1}
)$ means that
there are two error terms, one of which is bounded by
$C_1 \frac{1}{N\lambda^2}$ and the other by $C_2 \lambda^{M+1}$ with
$C_1, C_2$ depending on $f$ and $M$.
Especially, $C_2$ may grow in $M$ so that we cannot deduce
uniqueness of the invariant
measure for small $\lambda$ this way (cf. Remark \ref{Remark1} below).

When $\mathcal{R}\neq\one$ or $\mathcal{P}\neq0$ further
assumptions are needed in
order to control the Birkhoff sums.
We assume that $\mathcal{R}$ and $\mathcal{P}$ generate commuting
compact groups,
that is, $\mathcal{R}\mathcal{P}\mathcal{R}^{-1} =
\operatorname{Ad}_\mathcal{R}(\mathcal{P})=\mathcal{P}$
and the closed Abelian Lie groups $\langle\mathcal{R}\rangle=
\overline{\{\mathcal{R}^k \dvtx k\in\ZZ\}}$ and
$\langle\mathcal{P}\rangle=\overline{\{ \exp(\lambda\mathcal{P})
\dvtx \lambda\in\RR)\}}$
are compact.
While $\langle\mathcal{P}\rangle$ is always connected, $\langle
\mathcal{R}\rangle$
can possibly be disconnected. However, there exists $K\in\NN$ such
that $\langle\mathcal{R}^K \rangle$ is connected.
By considering the suspended
Markov process $(y^n)_{n\in\NN}$ with $y_n=x_{Kn}$ corresponding to
the family
\[
\Tt_{\lambda,\sigma_1,\ldots,\sigma_{K}} =
\Tt_{\lambda,\sigma_{K}} \cdots\Tt_{\lambda,\sigma_1}
\]
for $(\sigma_1,\ldots,\sigma_{K})\in(\Sigma^{K}, \mathbf{p}^{K})$, one
can always assume that $\langle\mathcal{R}\rangle$ is connected
and we shall do so from now on.
Note that the product $\langle\mathcal{R}\rangle\langle\mathcal
{P}\rangle$
is also a compact, connected, Abelian subgroup of $\Gg$
which will be denoted by $\langle\mathcal{R}, \mathcal{P}\rangle$.
All these groups are tori in $\Gg$ and their dimensions are
$L_\mathcal{R}$, $L_\mathcal{P}$ and $L_{\mathcal{R},\mathcal{P}}$. Hence,
$\langle\mathcal{R}\rangle\cong\TM^{L_\mathcal{R}},
\langle\mathcal{P}\rangle\cong\TM^{L_\mathcal{P}}$ and
$\langle\mathcal{R}, \mathcal{P}\rangle\cong\TM^{L_{\mathcal
{R},\mathcal{P}}}$,
where $\TM^L=\RR^L / (2\pi\ZZ)^L$ is the $L$-dimensional torus.
The (chosen) isomorphisms shall be denoted by $R_\mathcal{R},
R_\mathcal{P}$ and
$R_{\mathcal{R},\mathcal{P}}$, respectively, for example,
$R_\mathcal{R}(\theta) \in\langle\mathcal{R}\rangle\subset
\operatorname{GL}(L,\CC) $
for $\theta=(\theta_1,\ldots,\theta_{L_\mathcal{R}})\in\TM
^{L_\mathcal{R}}$.

The isomorphism $R_\mathcal{R}$ directly leads to the Fourier
decomposition of
the function $\theta\in\TM^{L_\mathcal{R}}\mapsto f(R_\mathcal
{R}(\theta)\cdot x)$,
notably
%
%
\begin{equation}
\label{eq-Fourier}
f(R_\mathcal{R}(\theta)\cdot x) = \sum_{j\in\ZZ^{L_\mathcal
{R}}} f_j(x)
e^{\imath j\cdot\theta}
,
\end{equation}
where
\[
f_j(x)
=
\int_{\TM^{L_\mathcal{R}}} \frac{d\theta}{(2\pi
)^{L_\mathcal{R}}}
e^{-\imath j\cdot\theta} f \bigl( R_{\mathcal{R}}(\theta)\cdot
x\bigr),\qquad
j\cdot\theta= \sum_{l=1}^{L_{\mathcal{R}}}j_l\theta_l
.
\]
Similarly, the maps
$\theta\in\TM^{L_\mathcal{P}}\mapsto f(R_\mathcal{P}(\theta)\cdot
x)$ and
$\theta\in\TM^{L_{\mathcal{R},\mathcal{P}}}\mapsto f(R_{\mathcal
{R},\mathcal{P}}(\theta)\cdot x)$
lead to Fourier series.
%
%
\begin{defini}
A function $f \in C^\infty(\Mm)$ is said to consist
of only low frequencies w.r.t.
$\langle\mathcal{R}\rangle$ if the Fourier coefficients $f_j \in
C^\infty(\Mm)$
vanish for $j$ with norm
$\|j\|=\sum_{l=1}^{L_{\mathcal{R}}} |j_l|$ larger than some fixed integer
$J>0$. Similarly, $f$ is defined
to consist of only low frequencies w.r.t. $\langle\mathcal{P}\rangle
$ or
$\langle\mathcal{R},\mathcal{P}\rangle$.
\end{defini}
%

The following definitions are standard
(see \cite{KDM} for references).
%
%
\begin{defini}
Let us define $\hat\theta_\mathcal{R}\in\TM^{L_\mathcal{R}}$
by $R_\mathcal{R}(\hat\theta_\mathcal{R})=\mathcal{R}$ and
$\hat\theta_\mathcal{P}\in\RR^{L_\mathcal{P}}$ by
$R_\mathcal{P}(\lambda\hat\theta_\mathcal{P})=\exp(\lambda
\mathcal{P})$.
Then $\mathcal{R}$ is said to be a Diophantine rotation or simply
Diophantine if there is
some $s>1$ and some constant $C$ such that for any nonzero multi-index
$j\in\ZZ^{L_\mathcal{R}} \setminus\{0\}$ one has
\[
| e^{\imath j \cdot\hat\theta_\mathcal{R}} - 1 |
\geq
C \|j\|^{-s} .
\]
Similar, $\mathcal{P}$ is said to be Diophantine, or a Diophantine
generator of
a rotation,
if there is some $s>1$ and some constant $C$,
such that for any nonzero multi-index $j\in\ZZ^{L_\mathcal
{P}}\setminus\{0\}$
one has
\[
| j \cdot\hat\theta_\mathcal{P}|
\geq
C \|j\|^{-s} .
\]
\end{defini}
%

As final preparation before stating the result, let us
introduce the measure $\overline{\mathbf{p}}$ on the Lie
algebra $\ggm$ obtained from averaging the distribution $\mathbf{p}_1$ of
the lowest-order terms $\mathcal{P}_{1,\sigma}$ w.r.t. the Haar measure
$dR$ on the
compact group $\langle\mathcal{R}, \mathcal{P}\rangle$,
namely for any measurable set $\bbb\subset\ggm$,
\[
\overline{\mathbf{p}}(\bbb)
=
\int_{\langle\mathcal{R},\mathcal{P}\rangle} dR\,
\mathbf{p} (
\{\sigma\in\Sigma\dvtx R\mathcal{P}_{1,\sigma} R^{-1} \in\bbb\}
).
\]
%
%
\begin{theo}
\label{theo-main}
Let $\Tt_{\lambda,\sigma}$ be of the form (\ref{eq-T}) and
$x_n$ the associated Markov process on $\Mm$ as given in (\ref{eq-Xn}).
Denote the Lie algebra of $\langle\mathcal{R},\mathcal{P}\rangle$
by $\rrr$ and
let $\vvv=\operatorname{Lie} (\operatorname{supp}(\overline{\mathbf{p}}), \rrr
)$ be
the Lie subalgebra of $\ggm$ generated by
the support of $\overline{\mathbf{p}}$ and $\rrr$.
Suppose that the smallest subgroup $\Vv$ of $\Gg$ containing $\{\exp
(\lambda\mathcal{P}) \dvtx \mathcal{P}\in\vvv, \lambda\in[0,1]\}
$ acts transitively on
$\Mm$.
Further, suppose that $f \in C^\infty(\Mm)$ and one of the following
conditions hold:

\begin{longlist}
\item $\mathcal{R}$ and $\mathcal{P}$ are Diophantine and
$\Mm=\Kkk/ \Hhh$ where $\Kkk$ and $\Hhh\subset\Kkk$ are compact
Lie groups.

\item
$f$ consist of only low frequencies w.r.t. $\langle\mathcal
{R},\mathcal{P}\rangle$.
\end{longlist}

Then there is a $\mu$-almost surely positive
function $\rho_0\in C^\infty(\Mm)$
normalized w.r.t. the Riemannian volume measure $\mu$ on $\Mm$,
such that
%
%
\begin{equation}
\label{eq-Birkhoffresult}
I_{\lambda,N}(f)
=
\int_\Mm\mu(dx) \rho_0(x) f(x) +
\Oo\biggl(\frac{1}{N\lambda^2}, \lambda\biggr) ,
\end{equation}
where $\mu$ is the Riemannian volume measure on $\Mm$.
Moreover, the probability measure $\rho_0\mu$ is invariant under
the action of $\langle\mathcal{R}, \mathcal{P}\rangle$.
\end{theo}
%

The probability measures $\sum_{m=0}^M \lambda^m \rho_m \mu$ in
Theorem \ref{theo-main2} and $\rho_0\mu$ in Theorem \ref{theo-main}
can be interpreted as perturbative approximations of
invariant measures $\nu_\lambda$. In fact, integrating
(\ref{eq-Birkhoffresult2})
over the initial condition $x_0$ w.r.t. any
invariant measure $\nu_\lambda$ and then taking the limit
$N\to\infty$, shows that for any smooth function
%
%
\begin{equation}
\label{eq-expansionformula}
\int_\Mm\nu_\lambda({d}x) f(x)
=
\sum_{m=0}^M \lambda^m \int_\Mm\mu({d}x) \rho_m(x)
f(x) +
\Oo(\lambda^{M+1} ) .
\end{equation}
This means that
the invariant measure is unique in a perturbative sense and, moreover,
its unique approximations are
absolutely continuous with smooth density.
In fact, one obtains the following.
%
%
\begin{coro}
\label{coro-main}
Let the assumptions of Theorems \ref{theo-main2} or
\ref{theo-main} be fulfilled and $(\nu_\lambda)_{\lambda>0}$
be a family of invariant probability measures for the Markov processes
$x_n(\lambda)$.
Then
\[
\mathrm{w}^*\mbox{-}\lim_{\lambda\to0} \nu_\lambda= \rho_0
\mu,
\]
where $\mathrm{w}^*\mbox{-}\lim$ denotes convergence in the weak-$*$
topology on the set of Borel measures.
\end{coro}
%
\begin{pf}
Approximating a continuous function by its Fourier series shows that
the set of smooth functions consisting of only low frequencies w.r.t.
$\langle\mathcal{R},\mathcal{P}\rangle$
is dense in the set of continuous functions w.r.t. the $\|\cdot\|
_\infty
$-norm.
The set of probability measures is norm bounded by $1$
w.r.t. the dual norm.
Now, let $g\in C(\Mm)$. For any $\varepsilon>0$, there is a smooth
function $g$ consisting of only low frequencies such that
$\|f-g\|_\infty< \varepsilon$. Then one has
\begin{eqnarray*}
|\nu_\lambda(f)-\rho_0\mu(f)| &\leq&
|\nu_\lambda(f-g)|+|\nu_\lambda(g)-\rho_0\mu(g)|+|
\rho_0\mu(g-f)| \\
&\leq&
2\varepsilon+|\nu_\lambda(g)-\rho_0\mu(g)| .
\end{eqnarray*}
One obtains ${\limsup_{\lambda\to0} }|\nu_\lambda(f)-\rho_0\mu
(f)| \leq
2\varepsilon$
for any $\varepsilon> 0$, so that by (\ref{eq-expansionformula})
\[
{\limsup_{\lambda\to0}} |\nu_\lambda(f)-\rho_0\mu(f)| = 0 ,
\]
for any continuous function $f \in C(\Mm)$, which gives the desired
result.
\end{pf}
\begin{Remark}\label{Remark1}
According to the unique weak-$*$-limit for a
family of invariant measures $\nu_\lambda$,
one might expect uniqueness for the invariant measure at least in a
small interval around $0$. However, we will briefly describe a
simple example satisfying
all conditions of Theorem \ref{theo-main2} such that for any rational
$\lambda$ the invariant measure is not unique.
Let $\Gg=\Mm={\mathbb S}^1=\{z\in\CC\dvtx |z|=1\}$ and let the Lie
group action
be the ordinary multiplication. Furthermore, let $\mathcal{R}=1$ and
$\mathcal{P}_{1,\sigma}$ be Bernoulli distributed with probability
$\frac12$ at
$\imath\pi$ and $-\imath\pi$ and let $\mathcal{P}_{n,\sigma}=0$ for
$n\geq2$. Any measure on ${\mathbb S}^1$ which is invariant
under a rotation by $\lambda\pi$ is an invariant measure and for
rational $\lambda$ there are many of them.
Therefore, we expect the following to hold:
given the conditions of Theorem \ref{theo-main} one finds
$\lambda_0>0$ such that for Lebesgue a.e. $\lambda\in[0,\lambda_0]$
there is a unique invariant measure.
\end{Remark}
\begin{Remark}\label{Remark2}
The main hypothesis of
Theorems \ref{theo-main2} and \ref{theo-main} is that the Lie group associated
to the Lie algebra $\vvv$ acts transitively on $\Mm$.
This can roughly be thought of as a Lie algebra equivalent of
Furstenberg's irreducibility condition
or the Goldsheid--Margulis criterion \cite{GM}.
Let us note that nontrivial
$\mathcal{R}, \mathcal{P}$ lead to a larger support for $\overline
{\mathbf{p}}$ and hence weaken
this hypothesis. A second hypothesis is that the group $\langle
\mathcal{R}\rangle$ is compact. This excludes many situations
appearing in
physical models where hyperbolic or parabolic channels appear. In
some particular situations, this could be dealt with \cite{SB1,SS}.
\end{Remark}
\begin{Remark}\label{Remark3}
As by the main hypothesis the action of
$\Gg$ on $\Mm$ is transitive,
$\Mm$~is always a homogeneous space
and given as a quotient of $\Gg$ w.r.t. some isotropy group, but
hypothesis (i) requires that $\Mm$ is, moreover, a quotient of a compact
group (which in the examples of Section \ref{sec-Lyap} is a
subgroup of $\Gg$).
The assumption that
$\Gg\subset\operatorname{GL}(L,\CC)$ (or, equivalently $\Gg$ has
a faithful representation) is only needed for the proof
of Theorem \ref{theo-main} under hypothesis (ii).
\end{Remark}
\begin{Remark}\label{Remark4}
Suppose $\Kkk$ is a compact subgroup of $\Gg$ acting transitively on
$\Mm$ [which is a special case of the condition in Theorem \ref{theo-main}(i)]. Then
the Haar measure $dk$ on $\Kkk$ induces a
unique natural $\Kkk$-invariant measure on $\Mm$ which one may choose
to be
$\mu$ (which is also the volume measure of the metric $\int dK\, K_*g$).
It is interesting to examine whether $\rho_0=1_\Mm$, that is,
the lowest-order approximation of the invariant measure is given by
the natural measure. The proof below provides a technique to check
this. More precisely, in the notation developed below,
$\hat{\Ll}^* 1_\Mm=0$ implies that $\rho_0$ is constant.
An example, where this can indeed be
checked is developed in Section \ref{sec-Lyap}.
Note that, if $\Kkk$ is as above, then any
conjugation $\mathcal{N}\Kkk\mathcal{N}^{-1}$ with an element
$\mathcal{N}\in\Gg$ has another
natural measure, given by $J_\mathcal{N}\mu$ where $J_\mathcal{N}$
is the Jacobian
of the map $x\mapsto\mathcal{N}\cdot x$. Unless $\mu$ is invariant
under all
of $\Gg$, the equality
$\rho_0=1_\Mm$ is hence linked to a good choice of $\Kkk$.
If $\mu$ is invariant under $\Gg$, then it is also an invariant
measure for the Markov process and under the hypothesis of
Theorem \ref{theo-main} one therefore has $\rho_0=1_\Mm$.
\end{Remark}
\begin{Remark}\label{Remark5}
If $\langle\mathcal{R},\mathcal{P}\rangle$ acts transitively on
$\Mm$, then
the measure $\rho_0 \mu$ is uniquely determined by the fact that it is
invariant under the action of $\langle\mathcal{R},\mathcal{P}\rangle
$ and normalized.
Moreover, $\Mm$ is isomorphic to the quotient of
$\langle\mathcal{R},\mathcal{P}\rangle$ and
the stabilizer $\Ss_x$ of any point $x\in\Mm$
(which is a compact Abelian subgroup of $\langle\mathcal{R},\mathcal
{P}\rangle$).
Hence, in this case, $\Mm$ is a torus and
the action is simply the translation on the torus. Consequently, the
measure $\rho_0 \mu$ is the Haar measure. Note that, if $\mathcal{P}=0$,
this holds independently of the perturbation and is imposed by the
deterministic process for $\lambda=0$.
\end{Remark}
\begin{Remark}\label{Remark6}
If the action of $\langle\mathcal{R}\rangle
$ on
$\Mm$ is not transitive, there are many
invariant measures $\nu_0$ for the deterministic dynamics (in
particular, if
$\mathcal{R}=\one$ any measure is invariant under $\langle\mathcal
{R}\rangle$).
Under the hypothesis of Theorems \ref{theo-main2} and
\ref{theo-main}, the random perturbations
$\mathcal{P}_{1,\sigma}$ and $\mathcal{P}_{2,\sigma}$ single out a
unique perturbative
invariant measure $\rho_0\mu$.
\end{Remark}
\begin{Remark}\label{Remark7}
We believe that condition that $\mathcal{R}$
and $\mathcal{P}$ commute is unnecessary. In fact, we expect that
conditions on $\mathcal{P}$ can be replaced by conditions on
$\hat\mathcal{P}=\int_{\langle\mathcal{R}\rangle} {d}R\,
R\mathcal{P}R^{-1}$.
\end{Remark}
\begin{Remark}\label{Remark8}
Let us cite prior work on the
rigorous perturbative evaluation of the averaged
Birkhoff sums (\ref{eq-avBirk}). In the case of $\Gg= $SL$(2,\RR)$,
$\Mm=\RR P(1)$ and a rotation matrix $\mathcal{R}$
in (\ref{eq-T}),
Pastur and Figotin \cite{PF}
showed (\ref{eq-Birkhoffresult}) for the lowest two
harmonics whenever $\mathcal{R},\mathcal{R}^2\neq\pm\one$.
The above result combined with the calculations in
Section \ref{sec-Lyap} shows that (\ref{eq-Birkhoffresult}) holds also
for other functions with $\rho_0=1_{\Mm}$. Without the
conditions $\mathcal{R},\mathcal{R}^2\neq\pm\one$,
Theorem \ref{theo-main} was proved in \cite{SB2,SS}.
Moreover, when $\mathcal{R}^K=\one$ (at so-called anomalies) and for
an absolutely
continuous distribution on~$\Gg$,
Theorem \ref{theo-main2} was proved by Campanino and Klein \cite{CK}.
Quasi-one-dimensional generalizations of \cite{PF} in the case where
$\Gg$ is a symplectic group were obtained in \cite{SB1,SB3}.
The work \cite{DP} is an attempt to treat higher-dimensional anomalies.
To further generalize, the above results to quasi-one-dimensional
systems was our main motivation for
this work.
\end{Remark}
\begin{Remark}\label{Remark9}
Our main application presented in
Section \ref{sec-Lyap}
is the perturbative calculation of Lyapunov
exponents associated to products of random matrices of the form
(\ref{eq-T}). Moreover, we show how to choose $\mathcal{N}$ (cf. Remark
\ref{Remark5}) such that $\rho_0=1_\Mm$.
This property is called the \textit{random phase
property} in \cite{RSB} which is related to the maximal entropy Ansatz
in the
physics literature.
Section \ref{sec-Lyap} can be read directly at this point
if Theorem \ref{theo-main} is accepted
without proof.
\end{Remark}
\begin{Remark}\label{Remark10}
The recent work by Dolgopyat and Krikorian \cite{DK} on random
diffeomorphisms on ${\mathbb S}^d$ contains results on the associated invariant
measure and Lyapunov spectrum which are related to the results of the
present paper. The main difference is that \cite{DK} assume the
random diffeomorphisms to be close to a set of rotations which
generate $\operatorname{SO}(d+1)$ while in the present work the diffeomorphisms
are to lowest order given by
the identity (Theorem 1) or close to one fixed rotation
(Theorem 2). As a result, the invariant measure in Proposition 2 of
\cite{DK} is close to the Haar measure while it is determined by the
random perturbations in the present paper. In the particular situation
of the example studied in Section \ref{sec-Lyap}, the randomness is such that the
invariant measure is the Haar measure and as a consequence the
Lyapunov spectrum is equidistant, just as in \cite{DK}.
\end{Remark}

In order to clearly exhibit the strategy of the proof of the
theorems, we first focus on the case $\mathcal{R}=\one$
and $\mathcal{P}=0$ in Sections \ref{sec-FP} and \ref{sec-PM}, which
corresponds to a higher-dimensional anomaly in the terminology of our prior work
\cite{SB2,SS}.
The main idea is then to expand $T_\lambda f$
into a Taylor expansion in $\lambda$. This directly leads to a
second-order differential operator $\Ll$ on $\Mm$ of the
Fokker--Planck type, for which the Birkhoff sums $I_{\lambda,N}(\Ll f)$
vanish up to order $\lambda$. Under the hypothesis of
Theorem \ref{theo-main2},
it can be shown to be a sub-elliptic H\"ormander operator on
the smooth functions on $\Mm$ with a one-dimensional cokernel.
Then one can deduce that $\CC+\Ll(C^\infty(\Mm))=C^\infty(\Mm)$ and
that the kernel of $\Ll^*$ is spanned by a smooth positive function
$\rho_0$. These are the main elements of the proof of
Theorem \ref{theo-main2} for $M=1$.
Then using the properties of the operators $\Ll$ and $\Ll^*$
and a further Taylor expansion of $T_\lambda f$ one can prove
Theorem \ref{theo-main2} by induction.
The additional difficulties for other $\mathcal{R}, \mathcal{P}$ in
Theorem \ref{theo-main}
are dealt with in the more technical Section \ref{sec-rotations}.
The applications to Lyapunov exponents are presented in
Section \ref{sec-Lyap}.

\section{Fokker--Planck operator and its properties}
\label{sec-FP}

In this section, we suppose $\mathcal{R}=\one$ and $\mathcal{P}=\EE
(\mathcal{P}_{1,\sigma})=0$
in (\ref{eq-T}) and introduce in this case the backward
Kolmogorov operator $\Ll$
and its adjoint $\Ll^*$, called forward Kolmogorov or also Fokker--Planck
operator \cite{Ris}. Their use for the calculation of the
averaged Birkhoff sum is exhibited and several properties of these
operators are studied. One
way to define the operator $\Ll\dvtx C^\infty(\Mm)\to
C^\infty(\Mm)$ is
%
%
\begin{equation}
\label{eq-FPdef2}
(\Ll f)(x) = \frac{{d}^2}{{d}\lambda^2} \bigg|_{\lambda
=0}(T_{\lambda}f)(x).
\end{equation}
Let us rewrite this using the smooth vector fields
$\partial_P$ associated to any
element $P\in\ggm$ by
%
%
\begin{equation}\label{eq-deltap}
\partial_P f(x)
= \frac{{d}}{{d}\lambda} \bigg|_{\lambda=0}
f(e^{\lambda P}\cdot x) .
\end{equation}
Then $\Ll$ is given by
%
%
\begin{equation}
\label{eq-FPdef}
\Ll=
\EE_\sigma(\partial_{\mathcal{P}_{1,\sigma}}^2 + 2
\partial_{\mathcal{P}
_{2,\sigma}} )
.
\end{equation}
%
%
\begin{proposi}
\label{prop-errorcontrol}
For $F\in C^\infty(\Mm)$, one has
\[
I_{\lambda,N}(\Ll F)
=
\Oo\biggl(\frac{1}{N\lambda^2} , \lambda\biggr) .
\]
\end{proposi}
%
\begin{pf}
For $P\in\ggm$, a Taylor expansion with Lagrange remainder gives
\[
F(e^P\cdot x) = F(x)+(\partial_P F)(x)+
\tfrac{1}{2} (\partial^2_P F)(x)+
\tfrac{1}{6} (\partial^3_P F)(e^{\chi P}\cdot x),
\]
for some $\chi\in[0,1]$. Choose $P=\lambda\mathcal{P}_{1,\sigma}+
\lambda^2\mathcal{P}_{2,\sigma}+\lambda^3 \Ss_\sigma(\lambda)$,
where $\Ss_\sigma(\lambda)=\sum_{n=3}^\infty\lambda^{n-3}\mathcal
{P}_{n,\sigma}$
and use that $\mathcal{P}_{1,\sigma}$ is centered
to obtain
\begin{eqnarray*}
\EE_\sigma F(\Tt_{\lambda,\sigma}\cdot x)
& = &
F(x) + \EE_\sigma\bigl( \lambda^2
\bigl(\tfrac{1}{2} \partial_{\mathcal{P}_{1,\sigma}}^2 F(x)+
\partial_{\mathcal{P}_{2,\sigma}}F(x) \bigr) \bigr)
+ \Oo(\lambda^3 ) \\
& = &
F(x) + \tfrac{1}{2} \lambda^2 \Ll F(x) +
\Oo(\lambda^3) .
\end{eqnarray*}
The error terms depend on derivatives of $F$ up to order 3
and are uniform in $x$ because $\Mm$ is compact and
$\mathcal{P}_{1,\sigma}, \mathcal{P}_{2,\sigma}$ and $\Ss_\sigma
(\lambda)$ are
compactly supported by (\ref{eq-taylorradius}).
Due to definition (\ref{eq-Xn}), this implies
\[
\EE_\omega\frac{1}{N} \sum_{n=1}^N F(x_n(\lambda,\omega))
=
\EE_\omega\frac{1}{N} \sum_{n=0}^{N-1}
F(x_n(\lambda,\omega)) + \frac{\lambda^2}{2} I_{\lambda,N}(\Ll
F) +
\Oo(\lambda^3) .
\]
As the appearing sums only differ by a boundary term, resolving
for $I_{\lambda,N}(\Ll F)$ finishes the proof.
\end{pf}

Next, let us bring the operator $\Ll$ into a normal form.
According to Appendix~\ref{appA}, one
can decompose $\mathcal{P}_{1,\sigma}$ into a finite linear
combination of
fixed Lie algebra vectors $\mathcal{P}_i\in\ggm$, $i\in I$,
with uncorrelated real random coefficients, namely
\[
\mathcal{P}_{1,\sigma}
=
\sum_{i=1}^{I}
v_{i,\sigma} \mathcal{P}_{i} ,\qquad
v_{i,\sigma} \in\RR,\qquad
\EE_\sigma(v_{i,\sigma})
= 0 ,\qquad
\EE_\sigma(v_{i,\sigma} v_{i',\sigma})
= \delta_{i,i'} .
\]
Then (\ref{eq-FPdef}) implies that $\Ll$ is in the so-called
H\"ormander form
\[
\Ll
=
\sum_{i=1}^I\partial_{\mathcal{P}_i}^2 + 2 \partial_{\Qq}
,
\]
where $\Qq=\EE_\sigma(\mathcal{P}_{2,\sigma})$.
Using the main assumption of Theorem \ref{theo-main2} (i.e.,
$\vvv\subset\uuu$),
one can show that $\Ll$ satisfies the strong H\"ormander property
of rank $r\in\NN$ \cite{Hor,RS,JS}.
%
%
\begin{proposi}
\label{prop-hoermander}
Under the assumptions of Theorem \ref{theo-main2},
there exists $r\in\NN$ such that
$\Ll$ satisfies a strong H\"ormander property of rank $r$,
that is, the vector fields
$\partial_{\mathcal{P}_i}$ and their $r$-fold commutators span the whole
tangent space at every point of $\Mm$.
\end{proposi}
%

In order to check this, one needs to calculate the
commutators of vector fields $\partial_P,\partial_Q$ for
$P,Q \in\ggm$. Let $X_P,X_Q$ denote the left-invariant vector fields
on $\Gg$ and furthermore introduce for each $x\in\Mm$ a function on
$\Gg$ by $f_x(\Tt)=f(\Tt\cdot x)$, $\Tt\in\Gg$. Then one obtains
\begin{eqnarray*}
\partial_P\partial_Q f(x)
&=&
\frac{{d}}{{d}\lambda} \bigg|_{\lambda=0}
(\partial_Q f)(e^{\lambda P}\cdot x)
=
\frac{{d}^2}{{d}\lambda\,{d}\mu} \bigg|_{\lambda,\mu
=0}
f(e^{\mu Q} e^{\lambda P}\cdot x)\\
&=&
X_Q X_P f_x(\one) ,
\end{eqnarray*}
which implies
%
%
\begin{eqnarray}
\label{eq-commutators}
(\partial_P \partial_Q - \partial_Q \partial_P) f(x)
&=&
(X_Q X_P - X_P X_Q) f_x(\one)
=
X_{[Q,P]} f_x(\one)\nonumber\\[-8pt]\\[-8pt]
&=&
\partial_{[Q,P]} f(x),\nonumber
\end{eqnarray}
where $[Q,P]$ denotes the Lie bracket (this is well known, see Theorem
II.3.4 in \cite{Hel}).
We also need the following lemma for the proof of
Proposition \ref{prop-hoermander}.
%
%
\begin{lemma}
\label{lem-Sard}
Let $\Uu\subset\Gg$ be a Lie subgroup of $\Gg$ that acts
transitively on $\Mm$ and denote the Lie algebra of $\Uu$ by $\uuu$.
Then the vector fields $\partial_P$, $P\in\uuu$,
span the whole tangent space at each point of $\Mm$.
\end{lemma}
%
\begin{pf}
First, let us show that there is a dense set of
points in $\Mm$ for which the vector fields $\partial_P$, $P\in\uuu$,
span the whole tangent space. Indeed, for a fixed
$x\in\Mm$ consider the surjective, smooth map
$\varphi_x \dvtx \Uu\to\Mm$, $\varphi_x(U)=U\cdot x$.
A point $x'\in\Mm$ is called regular for $\varphi_x$ if and only if
for any point in the preimage of $x'$ the differential $D\varphi_x$ is
surjective.
For each point $x'$, the hypothesis implies that there is
a $U\in\Uu$ such that $x'=\varphi_x(U)=U\cdot x$
and the regularity of $x'$ then shows that the paths $\lambda\mapsto
\varphi_x(e^{\lambda P}U)=e^{\lambda P}\cdot x'$, $P\in\uuu$, span
the whole tangent space at $x'$.
By Sard's theorem \cite{Hir}, the set of regular points is
dense in $\Mm$.

Actually, the existence of only 1 regular point $x$
implies that all points are regular. In fact, again
any other point is of the form $x'=U\cdot x$.
As the map $x\mapsto x'=U\cdot x$ is a diffeomorphism, the
push-forward of the paths $\lambda\mapsto\exp(\lambda P)\cdot x$,
$P\in\uuu$, given by the paths
$\lambda\mapsto U\exp(\lambda P)\cdot x=e^{\lambda UPU^{-1}}\cdot x'$,
$P\in\uuu$, span the tangent space also at $x'$.
\end{pf}
%
\begin{pf*}{Proof of Proposition \ref{prop-hoermander}}
Define iteratively the subspaces $\vvv_r \subset\ggm$ by
%
%
\begin{equation}
\label{eq-Liealgspaces}
\vvv_1
=
\operatorname{span}\{\mathcal{P}_i \dvtx  1\leq i \leq I\} ,\qquad
\vvv_r =
\operatorname{span} (\vvv_{r-1}\cup[\vvv_{r-1},\vvv_1] ) .
\end{equation}
By definition, one has
$\vvv_1=\operatorname{span}(\operatorname{supp}(\mathcal{P}_\sigma))$.
The space $\vvv\subset\ggm$ defined in Theorem~\ref{theo-main2}
is equal to $\vvv=\operatorname{Lie}(\vvv_1)$.
Due to (\ref{eq-commutators}), the strong H\"ormander property of
rank $r$ is equivalent to the property
that $\partial_P$, $P\in\vvv_r$, spans the whole
tangent space at every point $x\in\Mm$.

By the Lemma \ref{lem-Sard} and the assumption of
Theorem \ref{theo-main2}, this is fulfilled if
$\vvv_r =\vvv$ for some $r$.
As the vector spaces $\vvv_r$ are nested and $\ggm$ is finite dimensional,
the sequence has to become stationary. This means, there is some $r$
such that $\vvv_r=\vvv_{r+1}$. Using the Jacobi identity, one then checks
that $\vvv_r$ is closed under the Lie bracket and therefore $\vvv
_r=\vvv$.
\end{pf*}

Next, we want to recollect the consequences of the strong H\"ormander
property of rank $r$
as proved in \cite{Hor,RS,JS}. The first basic fact is the
subelliptic estimate within any chart
%
%
\begin{equation}
\label{eq-subelliptic}
\|f\|_{({1/r})}
\leq
C\bigl(\|\Ll f\|_{(0)}+\|f\|_{(0)}\bigr) ,
\end{equation}
where $\| \cdot \|_{(s)}$ denotes the Sobolev norms.
Using a finite atlas of $\Mm$, one can define a global
Sobolev space $H_s(\Mm)$ with norm also denoted by $\| \cdot \|_{(s)}$.
Then the estimate (\ref{eq-subelliptic}) holds also w.r.t. these
global norms.
Moreover, the norm $\| \cdot \|_{(0)}$ can be seen to be equivalent to
the norm in $L^2(\Mm,\mu)$ where $\mu$ is the Riemannian volume measure.
As usual, the embedding of $H_{s+\varepsilon}(\Mm)$ in $H_s(\Mm)$ is
compact for any $\varepsilon>0$.

The second basic fact is the hypoellipticity of $\Ll$. In order to
state this property, let us first extend $\Ll$ in the usual dual way
to an operator $\Ll_\dis$ on the space $\Dd'=(C^\infty(\Mm))'$
of distributions on $\Mm$. Then hypoellipticity states that,
for any smooth function $g$,
the solution $f$ of $\Ll_\dis f=g$ is itself smooth.

The Fokker--Planck operator $\Ll^*$ is the adjoint of $\Ll$ in
$L^2(\Mm,\mu)$. Because $\Mm$ is compact and has no boundary, the
domain $\Dd(\Ll^*)$ of $\Ll^*$ contains the smooth functions
$C^\infty(\Mm)$. Furthermore, $\Ll^*$ is again a second-order
differential operator with the same principal symbol as $\Ll$.
Therefore, $\Ll^*$ also satisfies the strong H\"ormander condition of
rank $r$. Thus, the subelliptic estimate as well as the
hypoellipticity property also holds for $\Ll^*_\dis$. We, moreover,
deduce that $\Ll$
is closable with closure $\overline{\Ll}=\Ll^{**}\subset\Ll_\dis$.

The following proposition recollects properties of $\Ll$ as a densely
defined operator on the Hilbert space $L^2(\Mm,\mu)$.
%
%
\begin{proposi}
\label{prop-dissipativ} There exists $c_0>0$ such that for $c>c_0$ the
following
holds:

\begin{longlist}
\item
$\Ll-c$ is dissipative.

\item $(\Ll-c)(C^\infty(\Mm))$ is dense in $L^2(\Mm
,\mu)$.

\item $\overline{\Ll}-c$ is maximally dissipative.

\item
$\overline{\Ll}-c$ is the generator of a contraction semigroup on
$L^2(\Mm,\mu)$.

\item The resolvent $(\overline{\Ll}-c)^{-1}$ exists and
is a
compact operator on $L^2(\Mm,\mu)$.
\end{longlist}
\end{proposi}
%
\begin{pf}
(i) Let us rewrite $\Ll$:
\[
\Ll f =
\sum_{i=1}^I [ \operatorname{div}(\partial_{\mathcal{P}_i}(f)
\partial_{\mathcal{P}_i}) - \operatorname{div}(\partial_{\mathcal
{P}_i})
\partial_{\mathcal{P}_i}(f) ] + 2\partial_Q(f).
\]
Defining
$X$ to be the smooth vector field
$2 \partial_\Qq-\sum_i\operatorname{div}(\partial_{\mathcal{P}_i})\partial
_{\mathcal{P}_i}$, one
has
\[
\Ll f =
\sum_{i=1}^I \operatorname{div}(\partial_{\mathcal{P}_i}(f)\,
\partial_{\mathcal{P}_i}) + X(f) .
\]
For a real, smooth function $f$, the divergence theorem and estimate
on the negative quadratic term gives
\[
\langle f | \Ll f\rangle
=
\int_\Mm{d}\mu\,
\Biggl[-\sum_{i=1}^I \partial_{\mathcal{P}_i}(f) \,\partial_{\mathcal
{P}_i}(f)
+ f X(f) \Biggr]
\leq
\int_\Mm{d}\mu \,f X(f)
.
\]
Using $2fX(f)=X(f^2)=\operatorname{div}(f^2X)-f^2\operatorname{div}(X)$ and
again the divergence theorem, it follows that
%
%
\begin{equation}
\label{eq-dissip}
\langle f | \Ll f\rangle
\leq
- \frac{1}{2} \int_\Mm{d}\mu\,\operatorname{div}(X) f^2
\leq \frac{1}{2} \|\operatorname{div}(X) \|_{\infty} \|f\|^2_2 .
\end{equation}
As $\Ll$ is real, it follows that
$\Re e\langle f | (\Ll-c) f\rangle\leq0$ for $f\in C^\infty(\Mm)$
and $c>c_0$ where $c_0=\frac{1}{2}\|\operatorname{div}(X) \|_{\infty}$.
By definition, this means precisely that $\Ll-c$ is dissipative.

(ii) Let $h\in L^2(\Mm,\mu)$ such that
$\langle h | \Ll f -cf\rangle= 0$
for all $f\in C^\infty(\Mm)=\Dd(\Ll)$.
Then $h$ is in the kernel of $\Ll^*_\dis$. By hypoellipticity,
it follows that $h\in C^\infty(\Mm)$.
Therefore, $\langle h | \Ll h \rangle= c \|h\|_2^2$ contradicting
(\ref{eq-dissip}) unless $h=0$.

The statement (iii) means that there is no dissipative
extension, which follows directly from (i) and (ii) by \cite{Dav}, Theorems 2.24,
2.25 and 6.4. Item (iv) follows from the same reference.

Concerning (v), the existence of the resolvent follows directly upon
integration of the contraction semigroup. Its compactness
follows from the subelliptic estimate (\ref{eq-subelliptic}) and
the compact embedding of $H_s(\Mm)$ into $L^2(\Mm,\mu)$.
\end{pf}

The next proposition is based on Bony's maximum principle for strong
H\"ormander operators \cite{Bon}, as well as standard Fredholm theory.
%
%
\begin{proposi}
\label{prop-kernel}
\textup{(i)} The kernel of $\overline{\Ll}$ consists of the
constant functions on $\Mm$.

{\smallskipamount=0pt
\begin{longlist}
\item[(ii)] The kernel\vspace*{1pt} of ${\Ll}^*$ is one dimensional and
spanned by a smooth function $\rho_0$.

\item[(iii)] $\operatorname{Ran} \overline{\Ll}=(\ker\Ll^*)^\perp
$ and
$\operatorname{Ran} \Ll^*=(\ker\overline{\Ll})^\perp=(\ker\Ll)^\perp$.

\item[(iv)] $\rho_0$ is $\mu$-almost surely positive.
\end{longlist}}
\end{proposi}
%
\begin{pf}
(i) By Corollaire 3.1 of \cite{Bon}, a smooth function $f$ which has a
local maximum and for which $\Ll f =0$ has to be constant on (the
pathwise connected compact set) $\Mm$.
If $f$ lies in the kernel of the closure $\overline{\Ll}=\Ll^{**}$, then
$\Ll_\dis f = 0$.
As $\Ll$ is hypoelliptic, $f \in C^\infty(\Mm)$ and therefore $f$
is again constant.

(ii)
Choose $c>c_0$ as in Proposition \ref{prop-dissipativ} and
let $K=(\overline{\Ll}+c)^{-1}$. Then one has
\begin{eqnarray*}
&&\overline{\Ll} f = g
\quad\Leftrightarrow\quad
(\overline{\Ll}+c)f = cf+g
\quad\Leftrightarrow\quad
f = cKf + Kg\\
&&\hphantom{\overline{\Ll} f = g} \quad\Leftrightarrow\quad
(\one-cK)f = Kg ,
\end{eqnarray*}
and similarly $\Ll^* f = g \Leftrightarrow (\one-cK^*)f=K^*g$.
For $g=0$, this implies $\ker\Ll=\ker(\one-cK)$ and $\ker\Ll^*=
\ker(\one-\bar cK^*)$. By the Fredholm alternative (the index of
$\one
+cK$ is $0$), the dimension of these two kernels are equal and by (i)
hence, both one dimensional. The smoothness of the function in the
kernel follows from the hypoellipticity of $\Ll^*$.

(iii)
For $ v\in\ker\Ll^*=\ker(\one- cK^*)$ and $\langle g | v\rangle=0$,
one has
$0=\langle g | v\rangle=\langle g | c K^*v \rangle=c \langle
Kg | v\rangle$,
therefore $g\in(\ker\Ll^*)^\perp$ implies $Kg \in
\ker(\one-cK)^\perp$ and the
Fredholm alternative states that $(\one-cK)f = Kg$ is solvable. Hence
by the
above, $\overline{\Ll} f = g$ is solvable. Therefore,
$\operatorname{Ran} \overline{\Ll}=(\ker\Ll^*)^\perp$. The other equality is
proved analogously.

(iv) Let $f\geq0$ be smooth
and suppose that $\int{d}\mu\,\rho_0 f=0$. According to (ii), (iii)
and hypoellipticity this implies that $f=\Ll F\geq0$ for some smooth
$F$. Again by Bony's maximum principle $F$ is constant and therefore
$f=0$. Hence, for any nonvanishing positive function $f$ one has
$\int{d}\mu\,\rho_0 f>0$.
\end{pf}

Even though not relevant for the sequel, let us also prove the following.
%
%
\begin{proposi}
\label{prop-Feller} $\Ll$ generates a
contraction semigroup in $(C(\Mm),\| \cdot \|_\infty)$, also called a
Feller semigroup.
\end{proposi}
%
\begin{pf}
This will follow directly from the Hille--Yosida theorem
\cite{Kal}, Theorem 19.11, once we verified that $(\Ll-c)C^\infty(\Mm)$ is
dense in $C(\Mm)$ for some $c>0$
and that $\Ll$ satisfies the positive-maximum
principle. The first property follows from the existence of the
resolvent (Proposition \ref{prop-dissipativ}) and the
hypoellipticity. For the second, let a smooth $f$ have a positive local
maximum at some $x\in\Mm$. Then one only has to check $(\Ll f)(x)\leq0$,
which follows because the first derivatives of $f$ vanish, its
second derivative is negative and the principal
symbol is positive definite.
\end{pf}

One can rewrite (\ref{eq-FPdef2}) as
$\lim_{\lambda\to0}\frac{1}{2\lambda^2}(T_\lambda-\one)f=\Ll f$ in
\mbox{$\| \cdot \|_\infty$} and for $f\in C^\infty(\Mm)$. Hence,
the above statement and \cite{Kal}, 19.28, implies
directly the following approximation result of the Feller process by
the discrete time
Markov processes.
%
%
\begin{coro}
\label{coro-approx} Let $e^{t\Ll}$ denote the Feller semigroup of
Proposition \ref{prop-Feller}. Then with convergence in
$(C(\Mm),\| \cdot \|_\infty)$,
\[
\lim_{\lambda\to0} T^{ [{t}/({2\lambda^2})]}_\lambda f
= e^{t\Ll}f .
\]
\end{coro}
%

Finally, let us note yet another representation of the
generator $\Ll$ following from the two above, namely
$\Ll=\lim_{N\to\infty} \frac{1}{2} N^\beta
((T_{N^{-\alpha}})^N-\one)$ where $\beta=2\alpha-1>0$ and
with strong convergence.

\section{Control of Birkhoff sum in the case $\mathcal{R}=\one,
\mathcal{P}=0$}
\label{sec-PM}

The aim of this section is the proof of
Theorem \ref{theo-main2}.
%
%
\begin{proposi}
\label{prop-invmeasure}
Let $\mathcal{R}=\one$ and $\mathcal{P}=0$.
The kernel of $\Ll^*_\dis$ is spanned by a nonnegative
smooth function $\rho_0$ that
is normalized by $\int_\Mm{d}\mu\, \rho_0= 1$.
For $f\in C^\infty(\Mm)$,
\[
I_{\lambda,N}(f)
=
\int_\Mm{d}\mu\,\rho_0 f +
\Oo\biggl(\frac{1}{N\lambda^2} , \lambda\biggr) .
\]
\end{proposi}
%
\begin{pf}
By hypoellipticity, the
kernel of $\Ll^*_\dis$ coincides with the kernel of~$\Ll^*$.
%
%
First, we show $C^\infty(\Mm)=\CC1_\Mm+ \Ll
C^\infty(\Mm)$.
Indeed,\vspace*{1pt} let $f\in C^\infty(\Mm)$. Set $C = \int_\Mm{d}\mu\, f
\rho
_0$ and
$\hat{f}=f-C$. Then one has $\int_\Mm{d}\mu\,\hat{f} \rho_0=0$
and therefore $\hat{f} \in(\ker\Ll^*)^\perp= \operatorname{Ran}\overline{\Ll
}$ by
Proposition \ref{prop-kernel}. By hypoellipticity, $\hat{f}\in\Ll
(C^\infty(\Mm))$.

Now using
Proposition \ref{prop-errorcontrol} and the above decomposition
\[
I_{\lambda,N} (f)
=
I_{\lambda,N}(\hat{f}+C)
=
C + I_{\lambda,N}(\Ll F)
=
C + \Oo(N^{-1}\lambda^{-2} , \lambda) ,
\]
one completes the proof.
\end{pf}

In order to prove Theorem \ref{theo-main2}, let us define the operators
\[
\Ll^{(M)} f(x)
=
\frac{{d}^{M}}{{d}\lambda^{M}}
\bigg|_{\lambda=0} (T_\lambda f)(x) ,\qquad
f \in C^\infty(\Mm) .
\]
Then $\Ll^{(1)}=0$ as $\mathcal{P}_{1,\sigma}$ is centered and $\Ll
^{(2)}=\Ll$.
Using (\ref{eq-T}), these operators can be written as
\[
\Ll^{(M)} f
=
\EE_\sigma\Biggl(
\sum_{m=0}^M \sum_{a_1+\cdots+a_m=M} \frac
{M!}{m!}
\partial_{\mathcal{P}_{a_1},\sigma}
\cdots \partial_{\mathcal{P}_{a_m},\sigma} f
\Biggr) .
\]
Hence, $\Ll^{(M)}$ is a differential operator of order $M$.
As $1_\Mm\in\ker\Ll^{(m)}$ and hence $\ker\Ll\subset\ker\Ll^{(m)}$
for all positive $m$, one obtains using Proposition \ref{prop-kernel}(iii)
\[
\operatorname{Ran} {\Ll^{(m)}}^* \subset
\bigl(\ker\Ll^{(m)}\bigr)^\perp\subset(\ker\Ll)^\perp
=\operatorname{Ran} \Ll^* .
\]
Therefore, and as $\ker\Ll^*$ is one dimensional, the functions
$\rho_M$ for $M\in\NN$ are iteratively and uniquely defined by
%
%
\begin{equation}
\label{eq-def-rho_m}
\Ll^* \rho_M
=
\sum_{m=1}^M \frac{2}{(m+2)!}{\Ll^{(m+2)}}^* \rho_{M-m} ,\qquad
\int_\Mm{d}\mu\,\rho_M=0,
\end{equation}
with $\rho_0$ given by Proposition \ref{prop-invmeasure}.
By induction and hypoellipticity of $\Ll^*$,
it follows that $\rho_M$ is a smooth function
for all $M$, therefore
the right-hand side of (\ref{eq-def-rho_m}) always exists. Now we can complete
the
following proof.
\begin{pf*}{Proof of Theorem \ref{theo-main2}}
The proof will be done by induction. The case $M=0$ is contained in
Proposition \ref{prop-invmeasure}.
For the step from $M-1$ to $M$, we first need a Taylor expansion of
higher order than
done so far. As $\mathcal{P}_{1,\sigma}$ is centered and due to the compact
support of $\mathcal{P}_{n,\sigma}$ and
$\sum_{m\geq n} \lambda^{m-n} \mathcal{P}_{n,\sigma}$ [uniform for small
$\lambda$ by (\ref{eq-taylorradius})],
one obtains with uniform error bound
\[
T_\lambda F (x)
=
F(x) + \frac{1}{2} \lambda^2 \Ll F(x) +
\sum_{m=3}^{M+2} \frac{\lambda^m}{m!} \Ll^{(m)} F(x)
+ \Oo(\lambda^{M+3}) ,
\]
which using the induction hypothesis
implies for Birkhoff sums
\begin{eqnarray*}
I_{\lambda,N}(\Ll F)
&=&
\sum_{m=1}^{M} \frac{2\lambda^m}{(m+2)!} I_{\lambda,N}\bigl(\Ll
^{(m+2)} F\bigr)
+
\Oo\biggl(\lambda^{M+1},\frac{1}{\lambda^2 N} \biggr) \\
&=&
\sum_{m=1}^{M} \sum_{l=0}^{M-m} \frac{2\lambda^{l+m} }{(m+2)!}
\int{d}\mu\,\rho_l \Ll^{(m+2)} F +
\Oo\biggl(\lambda^{M+1},\frac{1}{\lambda^2 N} \biggr)\\
&=&
\sum_{m=1}^{M}
\sum_{l=1}^m \frac{2 \lambda^m}{(l+2)!}
\int{d}\mu\, \bigl( {\Ll^{(l+2)}}^* \rho_{m-l} \bigr) F
+ \Oo\biggl(\lambda^{M+1},\frac{1}{\lambda^2 N} \biggr)\\
&=&
\sum_{m=1}^{M}\lambda^m \int{d}\mu\,\rho_m (\Ll F) +
\Oo\biggl(\lambda^{M+1},\frac{1}{\lambda^2 N} \biggr) .
\end{eqnarray*}
The last step follows from the definition
(\ref{eq-def-rho_m}) of $\rho_m$.
Now given any smooth function $f$, we can write it as
$f=\int{d}\mu\,\rho_0 f+\Ll F$ and obtain
\begin{eqnarray*}
I_{\lambda,N}(f)
&=&
\int{d}\mu\,\rho_0 f +
\sum_{m=1}^{M}\lambda^m \int{d}\mu\,\rho_m \Ll F
+ \Oo\biggl(\lambda^{M+1},\frac{1}{\lambda^2 N} \biggr)\\
&=&
\sum_{m=0}^{M}\lambda^m \int{d}\mu\,\rho_m f +
\Oo\biggl(\lambda^{M+1},\frac{1}{\lambda^2 N} \biggr),
\end{eqnarray*}
where the last step follows from $\int{d}\mu\,\rho_m =0$ for
$m\geq1$.
\end{pf*}

\section{Extension to lowest-order rotations}
\label{sec-rotations}

In this section, the lowest-order matrix $\mathcal{R}$ is
an arbitrary rotation and $\EE(\mathcal{P}_{1,\sigma})=\mathcal{P}$
commutes with $\mathcal{R}$ and
generates a rotation.
For any $R\in\Gg$, let us consider the
associated diffeomorphism $x\in\Mm\mapsto R\cdot x$ and its
differential $DR$. Then the push-forward
of functions $f\dvtx \Mm\to\CC$ and vector fields $X=(X_x)_{x\in\Mm}$ are
defined by
\[
(R_*f)(x) = f(R^{-1}\cdot x),\qquad
(R_* X)_{R\cdot x} = DR_x(X_x) .
\]
The pull-back is then $R^*=(R_*)^{-1}$. With this notation,
$R_*(X f)=(R_*X)(R_*f)$ and
\[
R_*(\partial_P(R^* f)) =
(R_*\partial_P)f =
\partial_{RPR^{-1}} f .
\]
Furthermore, we set $R_* (XY) = (R_* X)(R_* Y)$ for the composition of
two vector fields $X$ and $Y$.

Now let $\Ll$ be defined as in (\ref{eq-FPdef}) [note that this is
not equal
to the right-hand side of (\ref{eq-FPdef2})]. As $\mathcal{R}$ is a
zeroth-order term
in $\lambda$, the Birkhoff sums are to lowest order given by
averages along the orbits of $\mathcal{R}$.
Furthermore the expectation of the first-order term,
$\lambda\mathcal{P}$, then leads to averages over the group
$\langle\mathcal{P}\rangle$ to order $\lambda$.
It is hence
reasonable to expect that an averaged Kolmogorov operator has to be
considered.
In order to define it, recall that there are unique, normalized
Haar measures on the compact
groups $\langle\mathcal{R}\rangle$, $\langle\mathcal{P}\rangle$ and
$\langle\mathcal{R}, \mathcal{P}\rangle$.
Averages with respect to these measures
will be denoted by $\EE_{\langle\mathcal{R}\rangle}$,
$\EE_{\langle\mathcal{P}\rangle}$ and $\EE_{\langle\mathcal
{R},\mathcal{P}\rangle}$;
the integration variable
will be $R$. As the Haar measure is defined by left invariance and the groups
$\langle\mathcal{R}\rangle$ and $\langle\mathcal{P}\rangle$
commute by hypothesis, one has
$\EE_{\langle\mathcal{R},\mathcal{P}\rangle}(g(R)) = \EE
_{\langle\mathcal{P}\rangle}(\hat
{g}(R)$ for
$\hat{g}(\tilde{R})= \EE_{\langle\mathcal{R}\rangle} (g(\tilde
{R}R))$ and any
function $g$ on $\langle\mathcal{R},\mathcal{P}\rangle$.
Then set
%
%
\begin{equation}\label{eq-avFP}
\hat\Ll
=
\EE_{\langle\mathcal{R},\mathcal{P}\rangle} (R_* \Ll)
=
\EE_{\langle\mathcal{R},\mathcal{P}\rangle} \Biggl( \sum_{i=1}^I
\partial^2_{R\mathcal{P}_iR^{-1}} + \partial^2_{\mathcal{P}}
+ 2\partial_{R\Qq R^{-1}} \Biggr),
\end{equation}
where $\mathcal{P}_i$ are obtained by decomposing
the centered random variable $\mathcal{P}_{1,\sigma}-\mathcal{P}$
into a sum
$\sum_i v_{i,\sigma} \mathcal{P}_i $ such that the real coefficients
satisfy $\EE(v_{i,\sigma} v_{i',\sigma})=\delta_{i,i'}$ (cf.
Appendix \ref{appA}).
With this definition, we are able to prove a result similar to
Proposition~\ref{prop-errorcontrol}.
%
%
\begin{proposi}\label{prop-errorcontrol2}
Let $f\in C^\infty(\Mm)$ and
assume one of the following
conditions to hold:

\begin{longlist}
\item $\mathcal{R}$ and $\mathcal{P}$ are Diophantine
and $\Mm=\Kkk/ \Hhh$
for compact Lie groups $\Kkk$ and $\Hhh\subset\Kkk$.

\item $f$ consists of only low frequencies
w.r.t. $\langle\mathcal{R},\mathcal{P}\rangle$.
\end{longlist}

Then one has
\[
I_{\lambda,N}(\hat\Ll f)
=
\Oo\biggl(\frac{1}{N\lambda^2} , \lambda\biggr) .
\]
\end{proposi}
%

For the proof, we first need the following lemma.
%
%
\begin{lemma}\label{lemma-detmean0}
Let $f\in C^\infty(\Mm)$,
$f_0=\EE_{\langle\mathcal{R}\rangle}(R^*f)$ and
$\tilde{f}_0=\EE_{\langle\mathcal{P}\rangle}(R^*f_0)
=\EE_{\langle\mathcal{R},\mathcal{P}\rangle}({R}^*f)$.
If either \textup{(i)} or \textup{(ii)} as in the
Proposition \ref{prop-errorcontrol2} holds,
then
%
%
\begin{equation}
\label{eq-sum-for-estimate}
f-f_0 = g-\mathcal{R}^*g,\qquad f_0-\tilde{f}_0
=
\partial_\mathcal{P} \EE_{\langle\mathcal{R}\rangle} (R^* \tilde
g) ,
\end{equation}
for smooth functions $g,\tilde{g} \in C^\infty(\Mm)$.
\end{lemma}
%
\begin{pf}
The group $\langle\mathcal{R}\rangle$ is isomorphic to
a torus $\TM^{L_\mathcal{R}}$ with isomorphism
$R_\mathcal{R}(\theta)\in\langle\mathcal{R}\rangle$.
Furthermore we define $\hat\theta_\mathcal{R}$ by $\mathcal
{R}=R_\mathcal{R}(\hat\theta_\mathcal{R})$.
If $f$ consists of only low frequencies w.r.t.
$\langle\mathcal{R}, \mathcal{P}\rangle$, it can be
written as finite sum of its Fourier coefficients
\[
f
=
\sum_{\|j\|<J}
f_j \qquad \mbox{where }
f_j\bigl(R_\mathcal{R}(\theta)\cdot x\bigr) =
e^{\imath j\cdot\theta} f_j(x) ,
\]
where the Fourier coefficients are calculated as in
(\ref{eq-Fourier}).
Now set
\[
g
=
\sum_{0<\|j\|<J} \frac{f_j}{1-e^{\imath j\cdot\hat\theta
_\mathcal{R}}} .
\]
This is well defined because $\hat\theta_\mathcal{R}$ is irrational
as it
generates the whole torus.
Then $g-\mathcal{R}^* g=\sum_{0<\|j\|<J} f_j=f-f_0$.

As $\langle\mathcal{P}\rangle$ is an embedded subtorus
in $\langle\mathcal{R}, \mathcal{P}\rangle$, $f_0$ consists of only
low frequencies w.r.t. $\langle\mathcal{P}\rangle$. Let\vspace*{1pt}
$R_\mathcal{P}(\theta)$ denote the isomorphism of $\TM^{L_\mathcal
{P}}$ with
$\langle\mathcal{P}\rangle$ such that
$e^{\lambda\mathcal{P}}=R_\mathcal{P}(\lambda\hat\theta_\mathcal{P})$.
One can decompose $f_0=\EE_{\langle\mathcal{R}\rangle}((R)^*f)$
into a
Fourier sum w.r.t. the group $\langle\mathcal{P}\rangle$:
\[
f_0
=
\sum_{\|j\|<J}
\tilde{f}_j \qquad \mbox{where }
\tilde{f}_j\bigl(R_\mathcal{P}(\theta)\cdot x\bigr) =
e^{\imath j\cdot\theta} \tilde{f}_j(x) .
\]
Then
\[
\tilde{g} = \sum_{0<\|j\|<J}
\frac{\tilde{f}_j}{\imath j\cdot\hat\theta_\mathcal{P}}
\]
satisfies $\partial_\mathcal{P}\tilde{g}=f_0-\tilde{f}_0$. Furthermore,
$f_0-\tilde{f}_0$ is invariant under $\mathcal{R}$ which commutes with
$\mathcal{P}$, thus
\[
f_0-\tilde{f}_0 = \EE_{\langle\mathcal{R}\rangle}(R^*\partial
_\mathcal{P}\tilde{g})
= \partial_\mathcal{P} \EE_{\langle\mathcal{R}\rangle
}(R^*\tilde{g})
.
\]

In case (i), $g$ and $\tilde{g}$ will be defined by the same
formulas, but with infinite sums.
Thus, we have to show that these sums are well defined and that they
define smooth functions on $\Mm$.
Let $p\dvtx \Kkk\to\Mm$ be the projection identifying $\Mm$ with
$\Kkk/ \Hhh$ and define the smooth class
function $F(K,\theta)=f(R_\mathcal{R}(\theta)\cdot p(K))$
on the compact Lie group $\Kkk\times\TM^{L_\mathcal{R}}$.
We want to compare the Fourier series (\ref{eq-Fourier})
of $f$ w.r.t. $\mathcal{R}$
with the Fourier series of $F$ as given by the Peter--Weyl theorem.
By Theorem \ref{theo-K-with-torus} in Appendix \ref{appB},
this Fourier series of $F$ is given by
\[
f\bigl(R_\mathcal{R}(\theta)\cdot p(K)\bigr)
=
F (K,\theta)
=
\sum_{a\in\Ww_+} \sum_{j\in\ZZ^{L_\mathcal{R}}} d(a) \Tr
(\Ff F(a,j)
\pi_{a}(K) ) e^{\imath j \cdot\theta},
\]
where $\Ww_+$ denotes the set of highest weight vectors of $\Kkk$,
$\pi_a\dvtx \Kkk\to U(d(a))$ is the $d(a)$-dimensional,
unitary representation of $\Kkk$ parameterized by $a$,
and $\Ff F(a,j)$ is a $d(a)\times d(a)$ matrix
given by
\[
\Ff F(a,j)
=
\int_{\Kkk} {d}K \int_{\TM^{L_\mathcal{R}}}
{d}\theta\, F(K,\theta) \pi_{a}(K^{-1})
e^{-\imath j\cdot\theta} .
\]
Here, ${d}\theta$ and ${d} K$ denote the normalized Haar measures.
Comparing this equation with (\ref{eq-Fourier}),
one obtains that the Fourier coefficients w.r.t.
$\langle\mathcal{R}\rangle$ satisfy
\[
f_j(p(K))
=
\sum_{a\in\Ww_+} d(a) \Tr(\Ff F(a,j) \pi_{a}(K) ) .
\]
Let $g_j(x)=(1-e^{\imath j \cdot\hat\theta_\mathcal{R}})^{-1} f_j$
for $\|j\|>0$.
The next aim is to verify
that the infinite sum $g= \sum_{\|j\|>0} g_j$
defines a smooth function on $\Mm$.

As $F$ is smooth,
the Fourier coefficients $\Ff F(a,j)$ are rapidly decreasing by
\cite{Su} or Theorem \ref{theo-Su} in Appendix \ref{appB}, meaning that
$ \lim_{\|(a,j)\| \to\infty} \|(a,j)\|^h \|\Ff F(a$, $j)\| = 0$
for any natural $h$.
Here, one may choose some norm for which $\|(a,j)\|\geq\|j\|$ and
$\|\Ff F(a,j)\|$ denotes the Hilbert--Schmidt norm.
As $\mathcal{R}$ is Diophantine,
$|e^{\imath j \cdot\hat\theta_\mathcal{R}}-1| \geq C
\|j\|^{-s} \geq C \|(a,j)\|^{-s}$
for some natural $s$ and the coefficients
$\Ff G(a,j)= (1-e^{\imath j \cdot\hat\theta_\mathcal{R}}
)^{-1} \Ff F (a,j)$
defined for $\|j\|>0$ are still rapidly decreasing.
Therefore,
\[
G (K,\theta)
=
\sum_{\|j\|>0} \sum_{a\in\Ww_+} d(a) \Tr(\Ff G(a,j)
\pi_{a}(K) ) e^{\imath j \cdot\theta}
=
\sum_{\|j\|>0} g_j(p(K)) e^{\imath j\cdot\theta}
\]
is a smooth function and the series converges absolutely and
uniformly by Theorem \ref{theo-Su}. Setting $\theta=0$,
this implies that $\sum_{\|j\|>0} g_j$ converges uniformly to a smooth function
$g$ on $\Mm$ satisfying $g-\mathcal{R}^* g = \sum_{\|j\|>0} f_j = f-f_0$.

As before, we write $f_0=\EE_{\langle\mathcal{R}\rangle} (R^* f)$
as sum of Fourier coefficients
w.r.t. $\langle\mathcal{P}\rangle$, so $f_0 = \sum_j \tilde{f}_j$, and
let $\tilde{g}_j=(\imath j\cdot\hat\theta_\mathcal{P})^{-1} \tilde{f}_j$
for $\|j\|>0$. Consider the function
$\tilde F (K,\theta)=f_0(R_\mathcal{P}(\theta)\cdot p(K))$ on $\Kkk
\times
\TM^{L_\mathcal{P}}$, just as above define the Fourier coefficients
$\Ff\tilde F(a,j)$ for $a \in\Ww_+, j\in\ZZ^{L_\mathcal{P}}$
and let $\Ff\tilde G(a,j)=
(\imath j \cdot\hat\theta_\mathcal{P})^{-1} \Ff\tilde F(a,j)$.
As $|j \cdot\hat\theta_\mathcal{P}| \geq C \|j\|^{-s} \geq
C \|(a,j)\|^{-s}$ the coefficients
$\Ff\tilde{G}(a,j)$ are rapidly decreasing, the series
\[
\tilde{G}(K,\theta)
=
\sum_{a\in\Ww_+} \sum_{j\in\ZZ^{L_\mathcal{P}}} d(a)
\Tr(\Ff\tilde G(a,j) \pi_a(K) )
e^{\imath j \cdot\theta}
=
\sum_{\|j\|>0} \tilde{g}_j(p(K)) e^{\imath j \cdot\theta}
\]
converges absolutely and $\tilde{G}$ is smooth.
Thus, $\tilde{g} = \sum_{\|j\|>0} \tilde{g}_j$ exists, is smooth and
\[
\partial_\mathcal{P}\tilde{g}
= \frac{{d}}{{d}\lambda} \bigg|_{\lambda=0}
\sum_{\|j\|>0} \tilde{g}_j e^{\imath\lambda j\cdot\hat\theta
_\mathcal{P}}
=
\sum_{\|j\|>0} \tilde{f}_j
= f_0 - \tilde{f}_0 .
\]
As $f_0-\tilde{f}_0$ is $\mathcal{R}$-invariant one obtains also
$\partial_\mathcal{P} \EE_{\langle\mathcal{R}\rangle} (R^*
\tilde{g})= f_0 -
\tilde{f}_0$.
\end{pf}
%
%
\begin{lemma}\label{lemma-detmean}
If either \textup{(i)} or
\textup{(ii)}, as in Proposition \ref{prop-errorcontrol2} holds,
one has
\[
I_{\lambda,N}(f)
=
I_{\lambda,N}\bigl(\EE_{\langle\mathcal{R},\mathcal{P}\rangle}(R^* f)\bigr)
+ \Oo\biggl(\lambda,\frac{1}{\lambda N} \biggr) .
\]
\end{lemma}
%
\begin{pf}
Similarly as in the proof of Proposition \ref{prop-errorcontrol},
a Taylor expansion gives
\[
\EE_\sigma F(\Tt_{\lambda,\sigma}\cdot x)
=
\mathcal{R}^* F(x) + \lambda\partial_\mathcal{P} \mathcal
{R}^* F(x)
+
\frac{\lambda^2}{2} \Ll\mathcal{R}^* F(x)
+ \Oo(\lambda^3) ,
\]
where the error term is uniform in $x$.
For Birkhoff sums, this implies
%
%
\begin{equation}\label{eq-expansion-compl}\quad
I_{\lambda,N}(F-\mathcal{R}^*F)
=
\lambda I_{\lambda,N}(\partial_\mathcal{P}\mathcal{R}^*F) +
\frac{\lambda^2}{2} I_{\lambda,N}(\Ll\mathcal{R}^*F)
+ \Oo\biggl(\lambda^3 , \frac{1}{N} \biggr) .
\end{equation}
Using this for $F=g$, it
therefore follows that
$I_{\lambda,N}(f-f_0) = I_{\lambda,N}(g-\mathcal{R}^*g) =
\Oo(\lambda, N^{-1} )$.
The function $F=\EE_{\langle\mathcal{R}\rangle}(R^*\tilde{g})$
is $\mathcal{R}^*$-invariant, so
that the left-hand side of (\ref{eq-expansion-compl}) vanishes, and it follows
that
\[
I_{\lambda,N}(
f_0-\tilde{f}_0) =
I_{\lambda,N}\bigl(\partial_\mathcal{P}\EE_{\langle\mathcal
{R}\rangle}(R^*\tilde{g})\bigr)
=
\Oo\biggl(\lambda, \frac{1}{\lambda N} \biggr) .
\]
Combining both estimates completes the proof.
\end{pf}

As an immediate consequence, one obtains the following.
%
%
\begin{coro}
\label{coro-average-1st-order}
The derivative $d R_{\mathcal{R},\mathcal{P}}$ of the isomorphism
$R_{\mathcal{R},\mathcal{P}}\dvtx  \TM
^{L_{\mathcal{R},\mathcal{P}}}$ gives
an isomorphism from $\imath\RR^{L_{\mathcal{R},\mathcal{P}}}$ to
the Lie algebra $\rrr
$ of $\langle\mathcal{R},\mathcal{P}\rangle$.
Let $\Qq_1,\ldots,\break\Qq_{L_{\mathcal{R},\mathcal{P}}}$ be the images
of the standard orthonormal basis.
Then one has $\exp(2\pi\times\break\Qq_i)=1$ and the $\Qq_i$ span $\rrr$.
If either \textup{(i)} or
\textup{(ii)} as in Proposition \ref{prop-errorcontrol2} holds,
one has
\[
I_{\lambda,N}(\partial_{\Qq_i}(f)) = \Oo\biggl(\lambda,\frac
{1}{\lambda N} \biggr)\qquad
\mbox{implying }
I_{\lambda,N} \Biggl(\sum_{i=1}^{L_{\mathcal{R},\mathcal{P}}}
\partial^2_{\Qq_i}(f)
\Biggr) = \Oo\biggl(\lambda,\frac{1}{\lambda N} \biggr) .
\]
\end{coro}
%
\begin{pf}
First, note that $\partial_{\Qq_i} f$ consists
of only low frequencies w.r.t. $\langle\mathcal{R},\mathcal
{P}\rangle$
whenever $f$ does. By Lemma \ref{lemma-detmean}, it is sufficient to prove
$\EE_{\langle\mathcal{R},\mathcal{P}\rangle}(R^* (\partial_{\Qq
_i}f ))=0$.
This can be easily checked to be true as $\int_0^1 dt \exp(2\pi t \Qq
_i)^* (\partial_{\Qq_i} f)=0$.
\end{pf}

The following lemma is only needed for the proof of
Theorem \ref{theo-main} under hypothesis (ii).
%
%
\begin{lemma}
\label{lemma-conjpoly}
For any Lie algebra element $P\in\ggm$, smooth function
$f$ on $\Mm$ and any $x\in\Mm$, the map $\langle\mathcal
{R},\mathcal{P}\rangle\to\CC$,
$R\mapsto\partial^i_{RPR^{-1}} f(x), i\in\NN$,
is a trigonometric polynomial on $\langle\mathcal{R},\mathcal
{P}\rangle$
with uniformly bounded coefficients and uniform degree in $x\in\Mm$
(depending on $i$ though).
This implies that the function
$\Ll(\EE_{\langle\mathcal{R},\mathcal{P}\rangle}(R^* f))$
consists of only low frequencies
w.r.t. $\langle\mathcal{R},\mathcal{P}\rangle$.
\end{lemma}
%
\begin{pf}
As stated above, $\langle\mathcal{R},\mathcal{P}\rangle
\subset\Gg\subset\operatorname{GL}(L,\CC)$
is isomorphic to $\TM^{L_{\mathcal{R},\mathcal{P}}}$ and
the isomorphism is denoted by $R_{\mathcal{R},\mathcal{P}}(\theta
)\in\langle\mathcal{R},\mathcal{P}
\rangle$.
Furthermore, this group lies in some
maximal torus of $\operatorname{GL}(L,\CC)$.
As all maximal tori are conjugate to each other, so that
by exchanging $\Gg$ with some conjugate subgroup in
$\operatorname{GL}(L,\CC)$ one may assume $\langle\mathcal{R},\mathcal
{P}\rangle$ to be diagonal,
that is, it consists of diagonal matrices $R(\theta)=
\operatorname{diag}(e^{\imath\varphi_1(\theta)},\ldots,e^{\imath\varphi
_L(\theta)})$.
Beneath the $\varphi_1(\theta),\ldots,\varphi_L(\theta)$
there are maximally $L_{\mathcal{R},\mathcal{P}}$ rationally
independent, and each
is a linear combination with integer coefficients of
$\theta_1,\ldots,\theta_{L_{\mathcal{R},\mathcal{P}}}$. Hence, any
trigonometric
polynomial in
$\varphi(\theta)$ is a trigonometric polynomial in $\theta$
(possibly of higher degree), that is a trigonometric polynomial on
$\langle\mathcal{R},\mathcal{P}\rangle$.

On $\ggm\subset\operatorname{gl}(L,\CC)$, consider the usual real scalar product
${\Re e \Tr(P^* Q)}=\Re e \sum_{a,b} \overline{ P_{ab}} Q_{ab}$,
where $P_{ab}$ denotes the entries of the matrix $P$.
Let $M=\dim_\RR(\ggm)$ and $B^1,\ldots,B^M \in\ggm$ be some
orthonormal basis for $\ggm$ w.r.t. this scalar product.
If $R=\operatorname{diag}(e^{\imath\varphi_1},\ldots,e^{\imath\varphi_L})
\in\langle\mathcal{R},\mathcal{P}\rangle$ and $P\in\ggm$, then
one has
\begin{eqnarray*}
R P R^{-1}
&=&
\sum_{m=1}^M
\sum_{a,b=1}^L
\Re e ( \overline{B^m_{ab}} (R P R^{-1})_{ab} )B^m
\\
&=&
\sum_{m=1}^M
\sum_{a,b=1}^L
\Re e \bigl( \overline{B^m_{ab}} P_{ab} e^{\imath(\varphi
_a-\varphi
_b)} \bigr) B^m ,
\end{eqnarray*}
and therefore
\[
\partial_{R P R^{-1}}^i f
=
\Biggl(\sum_{m=1}^M
\sum_{a,b=1}^L
\Re e \bigl( \overline{B^m_{ab}} P_{ab} e^{\imath(\varphi
_a-\varphi
_b)} \bigr)
\partial_{B^m} \Biggr)^i f
\]
is a trigonometric polynomial in $\varphi$.
Thus by definition of $\Ll$, the map $R\mapsto R_*(\Ll( R^* f))
= (R_*\Ll) f$ is a trigonometric polynomial on
$\langle\mathcal{R},\mathcal{P}\rangle$, and therefore also
$R\mapsto R_* (\Ll\hat f)$ for $\hat f= \EE_{\langle\mathcal
{R},\mathcal{P}\rangle}(R^*f)$.
But this means precisely that $\Ll\hat f$ consists of only low
frequencies w.r.t.
$\langle\mathcal{R},\mathcal{P}\rangle$.
\end{pf}
\begin{pf*}{Proof of Proposition \ref{prop-errorcontrol2}}
As $\hat\Ll=\EE_{\langle\mathcal{R},\mathcal{P}\rangle}(R_* \Ll
)$, it follows for
$R\in\langle\mathcal{R},\mathcal{P}\rangle$
that $(R_* \hat\Ll)f = \hat\Ll f= (R^* \hat\Ll) f$.
This implies $R^* (\hat\Ll f) = \hat\Ll(R^* f)$ and
$\EE_{\langle\mathcal{R},\mathcal{P}\rangle}(R^* (\hat\Ll f)) =
\hat\Ll(\EE_{\langle
\mathcal{R},\mathcal{P}\rangle}(R^* f))$.
Hence, the Fourier coefficients of $\hat\Ll f$ are given by
%
%
\begin{equation}
\label{eq-Lf_j}
(\hat\Ll f)_j
=
\hat\Ll(f_j) .
\end{equation}
Therefore,\vspace*{2pt} $\hat\Ll f$ consists of only low frequencies w.r.t.
$\langle\mathcal{R},\mathcal{P}\rangle$ whenever $f$ does.
Furthermore, one obtains for $\hat f=\EE_{\langle\mathcal
{R},\mathcal{P}\rangle}(R^*
f)$ the following equalities:
\[
\EE_{\langle\mathcal{R},\mathcal{P}\rangle}(R^*( \hat\Ll f))
=
\hat\Ll\hat f
=
\EE_{\langle\mathcal{R},\mathcal{P}\rangle} (R_* (\Ll( R^* \hat f)))
=
\EE_{\langle\mathcal{R},\mathcal{P}\rangle} (R_*( \Ll\hat f)).
\]
Now $\Ll\hat f$ consists of only low frequencies by
Lemma \ref{lemma-conjpoly}.

Thus, applying Lemma \ref{lemma-detmean} twice
[the hypothesis are given either by hypothesis (i) of
Proposition \ref{prop-errorcontrol2} or by (ii) and
Lemma \ref{lemma-conjpoly}]. One obtains
\[
I_{\lambda,N}(\hat\Ll f)
=
I_{\lambda,N}\bigl(\EE_{\langle\mathcal{R},\mathcal{P}\rangle}(R^*
(\hat\Ll f))\bigr) +
\Oo\biggl(\lambda, \frac{1}{\lambda N} \biggr)
=
I_{\lambda,N}(\Ll \hat f) +
\Oo\biggl(\lambda, \frac{1}{\lambda N} \biggr) .
\]
As $\mathcal{R}^* \hat f=\hat f$ and $\partial_\mathcal{P}\hat f=0$,
equation (\ref{eq-expansion-compl}) for $F=\hat{f}$ implies
\[
I_{\lambda,N}(\Ll \hat f)
=
\Oo\biggl(\frac{1}{\lambda^2 N} , \lambda\biggr) ,
\]
which combined with the above finishes the proof.
\end{pf*}

After these preparations, the
proof of Theorem \ref{theo-main} is analogous to the case $\mathcal
{R}=\one$.
\begin{pf*}{Proof of Theorem \ref{theo-main}}
Consider the Markov process on $\Mm$ induced by the random family
\[
\Tt_{\lambda,\hat\sigma}
=
\exp(\lambda\mathcal{P}_{1,\hat\sigma} + \lambda^2
\mathcal{P}_{2,\hat\sigma
} ),
\]
where $\hat\sigma=(\sigma,R,\alpha,\beta,i) \in\hat\Sigma=
\Sigma\times\langle\mathcal{R},\mathcal{P}\rangle\times\{-1,1\}
\times\{-1,1\}\times
\{1,\ldots,L_{\mathcal{R},\mathcal{P}}\}$ and
$\mathcal{P}_{1,\hat\sigma}=(R\mathcal{P}_{1,\sigma
}R^{-1}-\mathcal{P})+\alpha\mathcal{P}+\beta\Qq_i
,
\mathcal{P}_{2,\hat\sigma}= R \mathcal{P}_{2,\sigma} R^{-1}$. The
$\Qq_i$ are defined
as in Corollary \ref{coro-average-1st-order}.
$\hat\Sigma$ is equipped
with the probability measure $\mathbf{p}\times{d}R \times\frac{1}{2}
(\delta_{-1}+\delta_1) \times
\frac{1}{2} (\delta_{-1}+\delta_1) \times\frac{1}{L_{\mathcal
{R},\mathcal{P}}}(\delta
_1 + \cdots + \delta_{L_{\mathcal{R},\mathcal{P}}})$
where ${d}R$ denotes the Haar measure on $\langle\mathcal
{R},\mathcal{P}\rangle$.
Let us define $\tilde\Ll=\hat\Ll+\sum_{i=1}^{L_{\mathcal
{R},\mathcal{P}}} \partial_{\Qq_i}^2$.
As $\EE_{\hat\sigma}(\mathcal{P}_{1,\hat\sigma})=0$,
\[
\tilde\Ll
= \sum_{i=1}^{L_{\mathcal{R},\mathcal{P}}} \partial_{\Qq_i}^2
+
\EE_{\langle\mathcal{R},\mathcal{P}\rangle} \EE_\sigma
(\partial_{R\mathcal{P}_{1,\sigma}R^{-1}-\mathcal{P}}^2+
\partial^2_\mathcal{P}+2 \partial_{R\mathcal{P}_{2,\sigma}
R^{-1}} )
=
\EE_{\hat\sigma}
(\partial_{\mathcal{P}_{1,\hat\sigma} }^2+2
\partial_{\mathcal{P}_{2,\hat\sigma}
} )
\]
and $\operatorname{span}(\operatorname{supp}(\mathcal{P}_{1,\hat\sigma}))=
\operatorname{span}(\operatorname{supp}(\overline{\mathbf{p}}),\rrr)$,
this new process leads to the operator $\tilde\Ll=\hat\Ll+\sum
_{i=1}^{L_{\mathcal{R},\mathcal{P}}} \partial_{\Qq_i}^2$
instead of $\Ll$ and the whole analysis done for $\Ll$ in the case
$\mathcal{R}
=\one, \mathcal{P}=0$
is applicable to $\tilde\Ll$ now due to the hypothesis of
Theorem \ref{theo-main}.
In particular, $\tilde\Ll$ and $\tilde\Ll^*$ are hypoelliptic operators,
the kernel of $\tilde\Ll$ consists of the constant functions and
the kernel of $\tilde\Ll^*$ is one-dimensional and spanned by a
normalized, smooth
function $\rho_0\geq0$.
Furthermore, $C^\infty(\Mm)=\CC1_\Mm+\tilde\Ll C^\infty(\Mm)$
and hence
for any smooth function $f$ and $C=\int_\Mm{d}\mu\,\rho_0 f$,
there is a smooth function $g$ such that $f=C+\tilde{\Ll} g$.

Assume $f$ consists of only low frequencies, that is,
$f_j =0$ for $\|j\|>J$.
Then by (\ref{eq-Lf_j}) one obtains for frequencies $\|j\|>0$
that $f_j = (f-C)_j = \tilde\Ll g_j$ and hence
$\tilde\Ll g_j = 0$ for $\|j\| \geq J$.
Therefore $g_j$ is constant, which means $g_j=0$
as $\|j\|>J>0$ and $g$ consists of only low frequencies if $f$ does.
Hence, Proposition~\ref{prop-errorcontrol2} implies for
both cases (i) and (ii)
the first statement of Theorem \ref{theo-main}:
\[
I_{\lambda,N}(f)
=
C + I_{\lambda,N}(\tilde{\Ll} g)
=
C + \Oo\biggl(\lambda,
\frac{1}{\lambda^2 N} \biggr) .
\]

To see that the measure $\rho_0\mu$ is
$\langle\mathcal{R},\mathcal{P}\rangle$-invariant, let
again $f$ be any smooth function.
As mentioned above, there exists $g\in C^\infty(\Mm)$ and $C\in\CC$
such that $\tilde\Ll g = f-C$. For all $R\in\langle\mathcal
{R},\mathcal{P}\rangle$,
this implies $\tilde\Ll R^*g=R^*\tilde\Ll g =R^*f -C$ and hence
$f-R^*f = \tilde\Ll(g-R^*g)\in(\ker\tilde\Ll^*)^\perp$
which gives
\[
\int_\Mm{d}\mu\,\rho_0 (f - R^*f)
= 0 .
\]
This is precisely the stated invariance property of the measure $\rho
_0\mu$.
\end{pf*}

\section{An application to random Jacobi matrices}
\label{sec-Lyap}

\subsection{Randomly coupled wires}
\label{sec-wires}

Here, we consider a family $H_\lambda$ of random Jacobi matrices with
matrix entries of the form
\[
(H_\lambda\psi)_n
= - \psi_{n+1} - \psi_{n-1} + \lambda W_{\sigma_n} \psi
_n
,\qquad
\psi=(\psi_n)_{n\in\ZZ}\in(\CC^L)^{\times\ZZ} ,
\]
where the $(W_{\sigma_n})_{n\in\ZZ}$ are independently drawn from an
ensemble of Hermitian $L\times L$ matrices, for which all the
entries $W_{i,j}\in\CC$, $1\leq i<j\leq L$, and $W_{k,k}\in\RR$,
$1\leq k\leq L$, are independent and
centered random variables with variances satisfying
%
%
\begin{equation}
\label{eq-cond-V}
\EE(W_{i,j}^2) = 0 ,\qquad \EE(|W_{i,j}|^2) = 1 ,\qquad
\EE(W_{k,k}^2) = 1.
\end{equation}
This is equivalent to having
$\EE(W_{i,j}W_{k,l})=\delta_{i,l}\delta_{j,k}$. This
model is relevant for the quantum mechanical description of a
disordered wire, consisting of $L$ identical subwires (all
described by a one-dimensional discrete Laplacian) which
are pairwise
coupled by random hopping elements having random magnetic phases.
Moreover, within each wire there is a random potential of the
Anderson type. This is similar to a model considered by Wegner
\cite{Weg} and Dorokhov \cite{Dor}.
We are interested in the weak coupling limit of
small randomness. Next, we show how this model leads to a question
which fits the framework of the main theorems of this work.

For a given fixed energy $E\in(-2,2)$, the associated transfer
matrices \cite{BL,PF} are
\[
\hat{\Tt}^{E}_{\lambda,\sigma} =
\pmatrix{
\lambda
W_\sigma-E \one& -\one
\cr
\one& \nul}
.
\]
Let us introduce the symplectic form $\Jj$, the Lorentz form $\Gg$ and
the Cayley transformation $\mathcal{C}$ by
\[
\Jj= \pmatrix{0 & - \one\cr\one& 0},\qquad
\Gg= \pmatrix{\one& 0 \cr0 & - \one},\qquad
\mathcal{C} = \frac{1}{\sqrt{2}} \pmatrix{\one& -\imath\one
\cr
\one& \imath\one}.
\]
Then the transfer matrix $\hat{\Tt}^{E}_{\lambda,\sigma}$ is in the
Hermitian symplectic group, namely it satisfies $\hat{\Tt}^*\Jj\hat
{\Tt
}=\Jj$. Hence, its Cayley transform $\mathcal{C}\hat{\Tt
}^{E}_{\lambda,\sigma
}\mathcal{C}^*$ is in the generalized Lorentz group U$(L,L)$ of signature
$(L,L)$ consisting by definition of the complex $2L\times2L$ matrices
$\hat{\Tt}$ satisfying $\Tt^*\Gg\Tt=\Gg$.
As a first step, let bring the transfer matrix in its normal form (this
corresponds to a change of conjugation as in the
proof of Lemma \ref{lemma-conjpoly}). Setting $E = - 2 \cos(k)$ and
\[
\mathcal{N} =
\frac{1}{\sqrt{\sin(k)}} \pmatrix{\sin(k) \one& 0
\cr
- \cos(k) \one& \one}
,
\]
where $|E|<2, \sin(k) \neq0$,
it is a matter of computation to verify
\[
\Tt_{\lambda,\sigma} = \mathcal{C}\mathcal{N} \hat{\Tt
}^{E}_{\lambda,\sigma} \mathcal{N}
^{-1} \mathcal{C}^*
= \mathcal{R}_k e^{ \lambda\mathcal{P}_\sigma}
\in \mathrm{U}(L,L) ,
\]
where
%
%
\begin{equation}
\label{eq-def-R+P}
\mathcal{R}_k
=
\pmatrix{e^{-\imath k} \one& 0
\cr
0 & e^{\imath k} \one},\qquad
\mathcal{P}_\sigma
=
\frac{\imath}{2\sin(k)} \pmatrix{W_\sigma& W_\sigma
\cr
- W_\sigma& - W_\sigma}.
\end{equation}
Note that the group generated by $\mathcal{R}_k$ is a subgroup of the group
consisting of all $\mathcal{R}_\theta$ for $\theta\in\TM$.
Furthermore, $\EE(W_\sigma)=0$.

The group $\mathrm{U}(L,L)$ naturally acts on the Grassmanian
flag manifold $\Mm$ of $\Gg$-isotropic subspaces of $\CC^{2L}$
\cite{BL}. In order to describe the flag manifold, let us introduce
the set of isotropic frames
\[
\IM
=
\{ \Phi\in\operatorname{Mat}(2L\times L,\CC) \dvtx  \Phi^*\Phi=\one;,
\Phi
^* \Gg\Phi= 0 \} .
\]
One readily checks that each $\Phi\in\IM$ is of the from
$\Phi=2^{-{{1/2}}}{U\choose V}$ with $U,V\in$U$(L)$. Hence,
$\IM\cong\mathrm{U}(L)\times\mathrm{U}(L)$ and it has
a natural measure given by the product of the Haar measures.
The column vectors of $\Phi$ then generate a flag. Two isotropic
frames $\Phi_1$ and $\Phi_2$ span the same flag if and only if there
is an upper triangular $L\times L$ matrix $S$ such that $\Phi_1=\Phi_2
S$. Due to the above, $S$ is also unitary so that it has to be a
diagonal unitary. These diagonal unitaries can be identified with the
torus $\TM^L$ and thus $\IM$ is a $\TM^L$-cover of the flag manifold,
namely $\Mm=\IM/\TM^L=\mathrm{U}(L)\times\mathrm{U}(L)/\TM^L$. Consequently
$\Mm$ is a symmetric space and it also carries a natural measure
$\mu$. The group action of $\mathrm{U}(L,L)$ on $\mathrm{U}(L)\times\mathrm{U}(L)$ is given
%
%
\begin{equation}
\label{eq-flagaction}
\pmatrix{A & B \cr C & D}
\cdot
\pmatrix{U \cr V}
=
\pmatrix{AU + BV \cr CU + DV} S ,
\end{equation}
where $S$ is an upper triangular matrix such that $(AU+BV)S$ is
unitary; then automatically also $(CU+DV)S$ is unitary. This also defines
an action on the quotient $\Mm$ and one readily checks that $\mu$ is
invariant under the action of the subgroup $\mathrm{U}(L,L)\cap\mathrm{U}(2L)$.

Let us recall how the general framework of the \hyperref[sec1]{Introduction} is applied
in the present situation: the Lie group is $\Gg=\mathrm{U}(L,L)$ acting on
the compact flag manifold $\Mm$ by (\ref{eq-flagaction}); equation
(\ref{eq-def-R+P}) shows that
the rotation is $\mathcal{R}=\mathcal{R}_k$ and the random perturbation
$\mathcal{P}_{1,\sigma}=\mathcal{P}_\sigma$, while $\mathcal
{P}_{n,\sigma}=0$ for $n \geq2$.
Objects of interest are now the $L$ positive Lyapunov exponents
$\gamma_{l,\lambda}(E)$, $l=1,\ldots,L$ \cite{BL}. It can be shown
that
%
%
\begin{equation}\label{eq-Lyapdef}
\sum_{l=1}^p \gamma_{l,\lambda}(E)
=
\lim_{N\to\infty} \EE\frac{1}{N}
\sum_{n=1}^N f_{p,\lambda}(x_n)
=
\lim_{N\to\infty}
I_{\lambda,N}(f_{p,\lambda}) ,
\end{equation}
where $x_n$ is the Markoff process on the compact manifold $\Mm$ and
$f_{p,\lambda}$ will be defined next. Actually, we may also consider
the action on the cover $\IM$ and then $f_{p,\lambda}$ is a class
function, defined for
$\Phi=(\phi_1,\ldots,\Phi_L)\in\IM$ by
\begin{eqnarray*}
f_{p,\lambda}(\Phi)
&=&
\EE_\sigma\log
(\| \Tt_{\lambda,\sigma} \phi_1 \wedge\cdots\wedge
\Tt
_{\lambda,\sigma} \phi_p \|_{\Lambda^p \CC^{2L}}
)
\\
&=&
\EE_\sigma\det_p (
\one_{p\times L}\Phi^* \Tt_{\lambda,\sigma}^* \Tt_{\lambda
,\sigma} \Phi
\one_{L\times p}),
\end{eqnarray*}
for $1\leq p\leq L$, where $\one_{p\times L}=(\one,0)$ is a $p\times L$
matrix and $\one_{L\times p}=\one_{p\times L}^*$.
Hence, $\gamma_{l,\lambda}(E)$ are all given by a Birkhoff sum.
Applying Theorem \ref{theo-main}, one obtains the following.
%
%
\begin{proposi}\label{prop-lyap}
As long as $E=2\cos(k)\neq0$ and $|E|<2$, the lowest-order
approximation $\rho_0\mu$ of the invariant measure is the Haar measure
on $\Mm$, that is, $\rho_0=1$. The $p$th greatest Lyapunov exponent
$\gamma_p(E)$
is then given by
%
%
\begin{equation}
\label{eq-Lyap-result}
\gamma_p(E)
=
\lambda^2 \frac{1+2(L-p)}{8 \sin^2(k)} + \Oo(\lambda^3) .
\end{equation}
\end{proposi}
%

For $L=1$, (\ref{eq-Lyap-result}) is proved in \cite{PF}.
At the band center $E=0$, the methods below show that the lowest-order
invariant measure is not the Haar measure. In the case $L=1$, the
measure was explicitly calculated in \cite{SB2}. A formula similar to
(\ref{eq-Lyap-result}) was obtained in \cite{Dor}. It shows, in
particular, that the Lyapunov spectrum is equidistant. Distinctness of
the Lyapunov exponents can also be deduced from the
Goldscheid--Margulis criterion.
The first step of the
proof is to expand $f_{p,\lambda}$ w.r.t. $\lambda$ for any $p$. To
deal with the expectation values, the following identities are
useful.
%
%
\begin{lemma}
\label{lem-calc}
\label{lemma-av-sigma}
Let $P,Q \in\operatorname{Mat}(L,\CC)$. Then one has
\begin{eqnarray*}
\EE(W_\sigma) &=& 0 ,\qquad
\EE(W_\sigma^2) = L \one,\qquad
\EE(\Tr(PW_\sigma)\Tr(QW_\sigma)) = \Tr(PQ) ,
\\
\EE(W_\sigma P W_\sigma) &=& \Tr(P) \one,\qquad
\EE(W_\sigma Q \overline{W}_\sigma) = Q^t.
\end{eqnarray*}
\end{lemma}
%

Using this, some calculatory effort leads to
%
%
\begin{equation}\label{eq-lyap2}
f_{p,\lambda}(\Phi)
=
\frac{\lambda^2}{8 \sin^2(k)}
F_p(\Phi)
+ \Oo(\lambda^3),
\end{equation}
where, setting $\Phi=2^{-{1/2}}{U\choose V}$, the class function
$F_p$ is defined by
\begin{eqnarray*}
F_p(\Phi)
&=&
2Lp+L\Tr\bigl( \one_{L;p} (V^*U+U^*V)
\bigr)
+\tfrac{1}{2} [\Tr(\one_{L; p} (U^* V) ) ]^2\\
&&{} + \tfrac{1}{2} [\Tr(\one_{L;p} V^* U ) ]^2
-p^2 ,
\end{eqnarray*}
where $\one_{L;p}=\one_{L\times p} \one_{p\times L}$ is the projection
on the first $p$ entries in $\CC^L$. Therefore,
%
\begin{equation}
\label{eq-Lyap-pert}
\sum_{l=1}^p \gamma_{l,\lambda}
=
\frac{\lambda^2}{8\sin^2(k)} \lim_{N\to\infty}
I_{\lambda,N}(F_p) + \Oo(\lambda^3) .
\end{equation}
Note that $F$ is a
polynomial of second degree in the entries of $(U, V),$ and hence
consists of only
low frequencies w.r.t. to $\langle\mathcal{R}_k \rangle$
as $\mathcal{R}_\theta{U\choose V} = {e^{-\imath\theta} U\choose
e^{\imath
\theta} V}$.
Thus, in order to apply Theorem \ref{theo-main} we just need to check
the coupling hypothesis.

\subsection{\texorpdfstring{Verifying the coupling hypothesis for Theorem
\protect\ref{theo-main}}{Verifying the coupling hypothesis for Theorem 2}}

First, we introduce a connected, transitively acting subgroup $\Uu
\subset\Gg$ such that
the space $\vvv$ as defined in Theorem \ref{theo-main} fulfills
$\uuu\subset\vvv$, where $\uuu$ is the Lie-algebra of $\Uu$.
Then $\Uu$ is also a subgroup of the group $\Vv$ as defined in
Theorem \ref{theo-main} and $\Vv$ acts transitively
as required.
Set
\[
\Uu
=
\{ \diag(U,V) \dvtx  U , V \in\mathrm{U}(L) \mbox{ and } UV \in
\operatorname{SU}(L) \}
\subset
\mathrm{U}(L,L) .
\]
Its Lie algebra is given by
\[
\uuu=
\{ \diag(u , v ) \dvtx
u , v \in \mathrm{u}(L) , \Tr(u+v)=0 \} .
\]
Now the action of $\Uu$ via (\ref{eq-flagaction}) on $\IM$ is not
transitive, but it is indeed transitive on the quotient $\Mm=\IM/\TM^L$.
%
%
\begin{proposi}
The Lie algebra $\uuu$ is contained in the Lie algebra $\vvv$ generated
by the set $\{ \mathcal{R}\mathcal{P}\mathcal{R}^{-1} \dvtx
\mathcal{R}\in \langle\mathcal{R}_k\rangle,
\mathcal{P}\in\operatorname{supp}(\mathcal{P}_\sigma)\}$, where
$\mathcal{P}_\sigma$ is given in (\ref{eq-def-R+P}).
\end{proposi}
%
\begin{pf}
We obtain
\[
\mathcal{R}_k \mathcal{P}_\sigma
\mathcal{R}_k^{-1}
=
\frac{\imath}{2 \sin(k)}
\pmatrix{W_\sigma& e^{-2\imath k} W_\sigma
\cr
-e^{2\imath k} W_\sigma& -W_\sigma}.
\]
Hence,
\[
-2 \cos(2k) \mathcal{P}_\sigma+ \mathcal{R}_k \mathcal
{P}_\sigma\mathcal{R}_k^{-1}
+ \mathcal{R}_k^{-1} \mathcal{P}_\sigma\mathcal{R}_k
= \frac{1-\cos(2k)}{\sin(k)}
\pmatrix{\imath W_\sigma& \nul
\cr
\nul& -\imath W_\sigma}.
\]
Therefore,\vspace*{1pt} the space $\vvv$ contains all matrices
$\bigl({\imath W \atop\nul} \enskip{\nul\atop-\imath W}\bigr)$
where $W=W^*$.
The commutator of such two matrices is
$ \bigl({[\imath V,\imath W] \atop\nul} \enskip{\nul\atop[\imath
V,\imath W]} \bigr)$,
hence also\vspace*{1pt} obtained in $\vvv$.
As $\operatorname{su}(L)$ is a simple Lie-algebra and
$\imath V$ and $\imath W$ are arbitrary elements of $\mathrm{u}(L)$,
the commutators $[\imath V,\imath W]$ contain any element of
$\operatorname{su}(L)$.
Therefore, taking linear combinations of these terms shows that
$\uuu\subset\vvv$.
\end{pf}

Thus, Theorem \ref{theo-main} applies and equation
(\ref{eq-Lyap-result}) follows readily from (\ref{eq-Lyap-pert}) once
one has shown that $\rho_0 \mu$ is the Haar measure
on $\Mm=\mathrm{U}(L)\times\mathrm{U}(L)/\TM^L$ for $E\neq0$. Furnishing
$\Mm$ with a left invariant metric, the Haar measure is the volume
measure so that we have to show $\rho_0=C 1_\Mm$ with some
normalization constant $C$.
This is equivalent to verifying that $\hat{\Ll}^* 1_\Mm=0$.
Using $\partial_P^*=-\partial_P-\operatorname{div}(\partial_P)$ and the
special form
$\hat\Ll=\EE_{\langle\mathcal{R}\rangle}\EE_\sigma
(\partial^2_{R\mathcal{P}_\sigma R^{-1}})$ of the Fokker--Planck
operator in the
present situation, one gets
%
%
\begin{eqnarray} \label{eq-cond-vol}
\hat{\Ll}^* 1_\Mm
&=&
\EE_{\langle\mathcal{R}\rangle} \EE_\sigma
\bigl( \bigl(
[\partial_{R\mathcal{P}_{\sigma} R^{-1}}+\operatorname{div}(\partial
_{R\mathcal{P}_{\sigma} R^{-1}})
]^2
\bigr)1_\Mm\bigr)\nonumber\\[-8pt]\\[-8pt]
&=&
\EE_{\langle\mathcal{R}\rangle} \EE_\sigma\bigl(
\partial_{R\mathcal{P}_{\sigma} R^{-1}} (\operatorname{div}(\partial
_{R\mathcal{P}_{\sigma} R^{-1}}))
+(\operatorname{div}(\partial_{R\mathcal{P}_{\sigma} R^{-1}}))^2
\bigr) .\nonumber
\end{eqnarray}
In order to calculate this further, one needs a formula for the
divergence of a vector field $\partial_\mathcal{P}$, which is the
object of the
next section.

\subsection{Divergence of vector fields}

Let
\[
\mathcal{P}
=
\pmatrix{A & B \cr B^* & D}
\in\mathrm{u}(L,L) ,\qquad A^*=-A ,\qquad D^*=-D .
\]
The aim of this section is to calculate the divergence of the vector field
$\partial_\mathcal{P}$ on $\Mm$. It can be lifted to a vector field on
$\IM\cong\mathrm{U}(L)\times\mathrm{U}(L)$. At the point $(U,V)$,
$\partial
_\mathcal{P}$ is given by the path
%
%
\begin{eqnarray}
\label{eq-vectorfields}
\qquad t &\mapsto&
\bigl(U(\one+t[U^*AU+U^*BV+S]),
V(\one+t[V^*DU+V^*B^*U+S])\bigr)\nonumber\\[-8pt]\\[-8pt]
&&{}+\Oo(t^2) .\nonumber
\end{eqnarray}
The upper triangular matrix $S$ is determined by the fact that it
has reals on the diagonal such that
$U^*AU+U^*BV+S$ is in the Lie algebra $\mathrm{u}(L)$.
This leads to $S+S^*=-U^*BV-V^*B^*U$.
In order to calculate $S-S^*$, let us
define the following $\RR$-linear function on $\operatorname{Mat}(L,\CC)$,
%
%
\begin{equation} \label{eq-def-w}
w(A)
=
\sum_{j<k}
[ E_{j,k} (A+A^*)^t E_{j,k} - E_{k,j} (A+A^*)^t E_{k,j} ] ,
\end{equation}
where $E_{j,k}$ is the matrix with a one at position $j,k$ and a zero elsewhere.
One obtains $S-S^*=w(-U^*BV) =-w(U^*BV) \in\mathrm{u}(L)$.
Hence, the path defining $\partial_\mathcal{P}$ at $(U,V)$ as in
(\ref{eq-vectorfields}) is given by
\begin{eqnarray*}
\exp(t\mathcal{P}) \cdot (U,V)
&=&
\bigl(U\bigl(\one+t\bigl[U^*AU+\tfrac12 (U^*BV-V^*B^*U)-\tfrac12 w(U^*BV)\bigr]\bigr), \\
& &\hspace*{6pt}
V\bigl(\one+t\bigl[V^*DV+\tfrac12 (V^*B^*U-U^*BV)-\tfrac12 w(U^*BV)\bigr]\bigr)
\bigr) .
\end{eqnarray*}
Hence, we associate to the induced (lifted) vector field the function
$P(U,V)=(U^*AU+\frac12 (U^*BV-V^*BU)-\frac12 w(U^*BV),
U^*DU+\frac12 (V^*B^*U-U^*BV)-\frac12 w(U^*BV)])$.

This vector field induces a projected vector field $\partial_\mathcal
{P}$ on
$\Mm$ and we want to
calculate its divergence on $\Mm$. The natural metric on $\mathrm{u}(L)\times\mathrm{u}(L)$
induced by the Killing form on $\mathrm{u}(2L)$ is given by
$\langle(u,v)|(\tilde{u},\tilde{v})\rangle= \Tr(u^*\tilde{u}+v^*
\tilde{v})$.
The Lie algebra $\mathfrak{h}$ of $\Hhh$ consists of the elements
$(\imath\Phi,\imath\Phi)$ for diagonal, real matrices $\Phi$.
An orthonormal basis
$(u_i,v_i)$ for $\mathfrak{h}^\perp$ in $\mathrm{u}(L)\times\mathrm{u}(L)$
is given
by the matrices
$\frac{1}{\sqrt{2}} (E_{j,k}-E_{k,j},\nul)$,
$\imath\frac{1}{\sqrt{2}} (E_{j,k}+E_{k,j},\nul)$,
$\frac{1}{\sqrt{2}} (\nul, E_{j,k}-E_{k,j})$,
$\imath\frac{1}{\sqrt{2}} (\nul, E_{j,k}+E_{k,j})$ and
$\imath\frac{1}{\sqrt{2}} (E_{j,j},- E_{j,j})$ for $1\leq j < k \leq L$.
The derivative w.r.t. to the left-invariant vector field on
$\mathrm{U}(L)\times\mathrm{U}(L)$ defined by $(u_i,v_i)$ will be denoted
by $\delta_{(u_i,v_i)}$.
According to (\ref{eq-div}) in Appendix \ref{appC} the divergence
$\operatorname{div}(\partial_\mathcal{P})$ on $\Mm$ is given by
\begin{eqnarray*}
&&\sum_i \delta_{(u_i,v_i)} \langle(u_i^*, v_i^*) | P(U,V) \rangle
\\
&&\qquad=
\sum_i \delta_{(u_i,v_i)} \biggl(
\Tr(u_i^* U^*AU +v_i^* V^*DV)
\\
&&\qquad\quad\hspace*{48.45pt}{}
-\frac{1}{2}
\Tr\biggl((u_i+v_i)^* w(U^*BV)\\
&&\qquad\quad\hspace*{48.45pt}\hspace*{37.36pt}{} +
\frac12 \Tr\bigl((u_i-v_i)^* (U^*BV-V^*B^*U)\bigr) \biggr)\biggr) .
\end{eqnarray*}
Now as $u_i^*=-u_i$, one obtains
\[
\delta_{(u_i,v_i)} \Tr(u_i^* U^*AU)
=
\Tr\bigl(u_i^*(u_i^* U^*AU+U^*AUu_i)\bigr)
=
\Tr\bigl(U^*AU(u_i^2-u_i^2)\bigr)
= 0 .
\]
Thus, one has $\sum_i \delta_{(u_i,v_i)}
\Tr(u_i^* U^*AU)=0$ and analogously $\sum_i \delta_{(u_i,v_i)}
\Tr(v_i^* V^*\times\break DV)=0$.
Next, consider $\sum_i \delta_{(u_i,v_i)} \Tr((u_i+v_i)w(U^*BV))$.
It is easy to check that for $j\neq k$ one has
$\sum_i u_i E_{j,k} \bar{v}_i = \sum_i \bar{u}_i E_{j,k} v_i= 0$ and
$\sum_i \bar{u}_i E_{j,k} u_i =
\sum_i u_i E_{j,k} \bar{u}_i = E_{k,j}$.
The same holds with $v_i$ and $u_i$ exchanged.
From these equations, the cyclicity of the trace and the definition of $w$
one obtains after some calculatory effort
\[
\frac{1}{2}\sum_i \delta_{(u_i,v_i)} \Tr\bigl((u_i+v_i)
w(U^*BV) \bigr)
=
\sum_{j<k} \Tr\bigl( (E_{k,k}-E_{j,j})(U^*BV+V^*B^*U) \bigr).
\]
The remaining term in $\operatorname{div}(\partial_\mathcal{P})$ is given by
\begin{eqnarray*}
&&\frac12 \sum_i \delta_{u_i,v_i} \Tr\bigl((u_i^*-v_i^*)(U^*BV-V^*B^*U)\bigr)
\\
&&\qquad=
\frac12 \sum_i \Tr\bigl((v_i-u_i)^2(U^*BV+V^*B^*U)\bigr) .
\end{eqnarray*}
As $\sum_i (v_i-u_i)^2 = -2L \one$, it follows that
%
\begin{equation}
\label{eq-div-dp}
\operatorname{div}(\partial_\mathcal{P})
=
2 \Re e \Tr(CU^*BV) ,
\end{equation}
where $C=-L\one+ \sum_{j<k} (E_{k,k}-E_{j,j})= \sum_{j=1}^L
(2j-1-2L) E_{j,j}$.
Note that $\operatorname{div}(\partial_\mathcal{P})$ is in fact a function on
$\Mm$,
that is, it is independent
on the choice of the preimage $(U,V)$ because $C$ is a diagonal matrix.

\subsection{Volume measure to lowest order}

For $E\neq0$, we now want to show $\hat\Ll1_\Mm=0$ using (\ref
{eq-cond-vol}).
%
%
As the group $\langle\mathcal{R}_k \rangle$ is a closed subgroup of the
torus consisting of all $\mathcal{R}_\theta$ for $\theta\in\TM=\RR
/2\pi\ZZ$, the
Haar measure of $\langle\mathcal{R}_k \rangle$ can be considered as a
probability measure on $\TM$. Expectations w.r.t. to this measure with
integration
variable $\theta\in\TM$ will be denoted by $\EE_\theta$.
Then for any function $f$ on $\langle\mathcal{R}_k \rangle$, one has
$\EE_\mathcal{R}(f(\mathcal{R}))=\EE_\theta(f(\mathcal{R}_\theta))$.
\begin{lemma}
\label{lemma-av-theta}
Away from the band center $E\neq0$, one has
\[
\EE_\theta(e^{\pm2 \imath\theta})
= 0 ,\qquad
\EE_\theta(e^{\pm4\imath\theta})
=
0 .
\]
\end{lemma}
\begin{pf}
If $k$ is an irrational angle, that is, $\frac{k}{2\pi}$
is irrational,
then the closed group generated by $\mathcal{R}_k$ is just the set of all
$\mathcal{R}_\theta$ and the measure $\EE_\theta$ is the Haar
measure of
the torus $\TM$ implying
$\EE_\theta(e^{\pm2\imath\theta})=\EE_\theta(e^{\pm4\imath
\theta})=0$.
If $k$ is a rational angle, then the closed group generated
by $\mathcal{R}_k$ is finite and consists of all $\mathcal{R}_\theta
$ such that
$e^{\imath\theta}$ is a $s$th root of $1$ for some natural $s$.
The Haar measure is just the point measure giving each point the same mass.
As $\sin(k) \neq0$, we get $s>2$ which gives $\EE_\theta(e^{\pm
2\imath\theta})=0$.
Similarly, as long as $s\neq4$ one also obtains $\EE(e^{\pm4\imath
\theta})=0$.
If $s=4$
which means $k= \pi/2$ and $E=0$, then
$\EE_\theta(e^{4\imath\theta})=1$.
\end{pf}

Define $A_\sigma= B_\sigma=\frac{\imath W_\sigma}{2 \sin^2(k)}$ and
$D_\sigma= -A_\sigma$. Then
\[
\mathcal{R}_\theta\mathcal{P}_\sigma
\mathcal{R}_{\theta}^{-1}=
\pmatrix{A_\sigma & e^{-2\imath\theta} B_\sigma\vspace*{2pt}\cr
e^{2\imath\theta} B^*_\sigma & D_\sigma}.
\]
From now on, we assume $E\neq0$. First, consider the term
$[\operatorname{div}(\partial_{\mathcal{R}_\theta\mathcal{P}_\sigma\mathcal
{R}_\theta^{-1}} )]^2$ appearing in (\ref{eq-cond-vol}). By
(\ref{eq-div-dp}), it is equal to
\[
e^{-4\imath\theta}\Tr(CU^*B_\sigma V)^2+
e^{4\imath\theta} \Tr(CV^*B^*_\sigma U)^2
+2\Tr(CU^*B_\sigma V) \Tr(CV^*B^*_\sigma U) .
\]
By Lemmas \ref{lemma-av-theta} and \ref{lemma-av-sigma},
one obtains
\[
\EE_\mathcal{R}\EE_\sigma
(\operatorname{div}(\partial_{\mathcal{R}\mathcal{P}_\sigma\mathcal
{R}^{-1}})(U,V) )^2
=
\frac{1}{2 \sin^2 k} \Tr(VCU^*UCV^*)
=
\frac{\Tr(C^2)}{2 \sin^2(k)}
.
\]
Next, we need to calculate the average of
$
\partial_{\mathcal{R}_\theta\mathcal{P}_\sigma\mathcal{R}_\theta
^{-1}} \operatorname{div}
(\partial_{\mathcal{R}_\theta\mathcal{P}_\sigma\mathcal{R}_\theta^{-1}})
$ which equals
\begin{eqnarray*}
& &
\Re e \Tr\bigl(e^{-2\imath\theta}2CU^*(A_\sigma^*B_\sigma
+B_\sigma
D_\sigma) V
+C(V^*B_\sigma^*B_\sigma V+U^*B_\sigma B_\sigma^*U) \\
& &\hspace*{40.02pt}{}
- e^{-4\imath\theta} 2U^*B_\sigma VU^*B_\sigma V -
e^{-2\imath\theta} U^*B_\sigma V
(Cw^*_{\theta,\sigma} +w_{\theta,\sigma}C ) \bigr) ,
\end{eqnarray*}
where $w_{\theta,\sigma}=w(e^{-2\imath\theta} U^*B_\sigma V)$ and $w$
is defined as in (\ref{eq-def-w}).
Averaging over $\langle\mathcal{R}_k \rangle$ and $\sigma$ one gets
by Lemma \ref{lemma-av-theta} and Lemma \ref{lemma-av-sigma}
that $\EE_\mathcal{R}\EE_\sigma(\partial_{\mathcal{R}\mathcal
{P}_\sigma\mathcal{R}^{-1}}\*\operatorname{div}
(\partial_{\mathcal{R}\mathcal{P}_\sigma\mathcal{R}^{-1}}))$ is
equal to
\[
\frac{L \Tr(C)}{2 \sin^2(k)}
- \EE_\theta\EE_\sigma\Re e \bigl(
e^{-2\imath\theta} \Tr\bigl(U^*B_\sigma V(Cw_{\theta,\sigma
}^*+w_{\theta
,\sigma}C)\bigr)
\bigr) .
\]
The last term with $w_{\theta,\sigma}$ consists of terms of the form
$e^{-4\imath\theta} \Tr(U^*B_\sigma VE_{k,j} (U^*\times\break BV)^t E_{k,j}C)$ and
$ \Tr(U^*B_\sigma VE_{j,k}U^t\overline{B}_\sigma\overline{V}
E_{j,k}C) $.
The latter one gives
$\frac{1}{4\sin^2(k)}\times\break\Tr(U^*U E_{k,j} V^t \overline{V} E_{j,k} C)=
\frac{1}{4 \sin^2(k)} \Tr(E_{k,k} C)$ after averaging over $\sigma$.
Therefore and by a similar result for the term with $w_{\theta,\sigma}^*$
as well as the definition of $C$, one obtains
\begin{eqnarray*}
\EE_\theta\EE_\sigma\Tr\bigl( e^{-2\imath\theta}U^*B_\sigma V
(w_{\theta,\sigma} C +Cw_{\theta,\sigma}^*) \bigr)
&=&
\frac{\sum_{j<k}\Tr((E_{k,k}-E_{j,j})C)}{2 \sin^2(k)}\\
&=&
\frac{\Tr((C+L\one)C)}{2\sin^2(k)} .
\end{eqnarray*}
Putting everything together one has
\[
\EE_\mathcal{R}\EE_\sigma(\operatorname{div} (\operatorname{div}(\partial
_{\mathcal{R}\mathcal{P}
_\sigma\mathcal{R}^{-1}})
\partial_{\mathcal{R}\mathcal{P}_\sigma\mathcal{R}^{-1}}
) )
=
\frac{\Tr(C^2)+L\Tr(C)-\Tr((C+L\one)C)}{2 \sin^2(k)}
=
0 .
\]
Therefore the lowest-order invariant measure $\rho_0 \mu$ on $\Mm$ is
given by the
Haar measure.

\begin{appendix}
\section{Vector-valued random variables}\label{appA}

%
\begin{lemma}
Let $a=(a_1,\ldots,a_n)^t\dvtx  \Sigma\to\RR^n$ be a centered,
vector-valued random variable on a probability space $(\Sigma,\mathbf{p})$,
and each $a_k \in L^2(\Sigma,\mathbf{p})$. Then there exist a linear
decomposition $a=\sum_i v_i b_i$ over finitely many
fixed vectors $b_i\in\RR^n$ with coefficient $v_i$ which
are centered random
variables $v_i \in L^2(\Sigma,\mathbf{p})$ that are uncorrelated
$\EE(v_i v_{i'})= \EE(v_i^2) \delta_{i,i'} $.
\end{lemma}
%
\begin{pf}
One can assume that the random variables $a_k$ as elements on
$L^2(\Sigma,\mathbf{p})$ are linearly independent [otherwise
one takes a basis for the vector space $\operatorname{span} (\operatorname{supp}(a))$
and rewrites the random variable $a$ as vector using this basis].
Let us introduce $\lambda_{k,j}$ for $k>j$ and write the Ansatz
$v_k=a_k+ \sum_{i=1}^{k-1} \lambda_{k,i} a_i$.
Inverting the matrix form of these equations gives
\[
\pmatrix{
a_1 \cr\vdots\cr a_n}
=
\pmatrix{1 & 0 & \cdots& 0 \cr
\lambda_{2,1} & 1 & \ddots& \vdots\cr
\vdots& \ddots& \ddots& 0 \cr
\lambda_{n,1} & \cdots& \lambda_{n,n-1} & 1}^{-1}
\pmatrix{v_1 \cr\vdots\cr v_n}.
\]
Hence, one can write $a$ as a sum $\sum_k v_k b_k$ where the $b_k$'s
are the vectors of the inverted matrix.
The $v_k$'s are pairwise uncorrelated,
if $\EE(v_k a_i)=0$ for all $i<k$, as this implies
$\EE(v_k v_i)=0$ for all $i<k$.
Now $\EE(v_k a_i)=0$ for $i=1,\ldots,k-1$ is guaranteed if
\[
{-} \pmatrix{
\EE(a_k a_1) \cr\vdots\cr\EE(a_k a_{k-1})}
=
\pmatrix{\EE(a_1 a_1) & \cdots& \EE(a_1 a_{k-1}) \cr
\vdots& \ddots& \vdots\cr
\EE(a_{k-1} a_1) & \cdots& \EE(a_{k-1} a_{k-1})}
\pmatrix{\lambda_{k,1} \cr\vdots\cr\lambda_{k,k-1}}.
\]
If the appearing matrix is invertible,
one can resolve this equation to get
$\lambda_{k,i}$ for all $i<k$.
So it remains to show that this matrix is invertible which is
equivalent to the property that the columns are linearly independent.
Now let $\xi_i \in\RR$ such that
\[
\sum_{i=1}^{k-1} \xi_i \EE(a_j a_i)
=
\EE\Biggl(a_j \sum_{i=1}^{k-1} \xi_i a_i \Biggr)
= 0
\]
for all $j=1,\ldots,k$. The vector
$\sum_{1\leq i\leq k-1} \xi_i a_i $ is then orthogonal in
$L^2(\Sigma,\mu)$ to any vector in the
subspace spanned by $a_1,\ldots,a_{k-1}$ and it
therefore has to be zero.
As the random variables $a_i$ are linearly
independent, one gets $\xi_i=0$ for all $i=1,\ldots,k-1$.\vspace*{-14pt}
\end{pf}

\section{Fourier series on compact Lie groups}\label{appB}

First, let us summarize some facts about the representation theory
of compact Lie groups. All this is well known and proofs can be found
in the literature, for example,
\cite{Bu}, but we need to introduce the
notation for the proof of Theorem \ref{theo-K-with-torus}.

Let $\Kkk$ be a compact Lie group equipped with its normalized Haar measure
and let $\Ttt\subset\Kkk$ be some maximal torus
$\Ttt\cong\TM^r$, where $r$ is called the rank of $\Kkk$.
The continuous irreducible representations of the torus $\Ttt$ are given
by the characters, that is,
the homomorphisms into the group $S^1=\mathrm{U}(1) \subset\CC$.
Let us denote them by $X^*(\Ttt)$.
They form a $\ZZ$-module isomorphic to the lattice $\ZZ^r$ and
hence $X^*(\Ttt)$ is a lattice in the vector
space $\Vv=\RR\otimes_\ZZ X^*(\Ttt)$, the tensor product over the
ring $\ZZ$.
This is an abstract description of the fact, that the characters of the torus
$\TM^r$ are given by the maps
$\theta\in\TM^r \mapsto e^{\imath j \cdot\theta}$ for a fixed $j
\in
\ZZ^r$.
In this case, $\Vv=\RR^r$.

Define some $\operatorname{Ad}_\Kkk$-invariant scalar product on the
Lie algebra $\mathfrak{k}$ of
$\Kkk$, where $\operatorname{Ad}_\Kkk$ denotes the adjoint representation, and adopt
$\Vv$ with an scalar product $\langle\cdot, \cdot\rangle$ such that
the norm of $a \in X^*(\Ttt)$ coincides with the operator norm of the
derivative ${d}a$ acting on $\mathfrak{t}$, the Lie algebra of
$\Ttt$.

Let $\ppp$ be the orthogonal complement in $\mathfrak{k}$
of $\mathfrak{t}$, the Lie algebra of $\Ttt$.
Then the group $\Ttt$ acts
on the complexification $\ppp_\CC= \CC\otimes_\RR\ppp$ by
the adjoint representation and linearity. This representation of $\Ttt$
can be decomposed into irreducible continuous representations,
which means $\ppp_\CC= \bigoplus_{a\in\Phi} \ppp_a$ where
$\ppp_a$ is the set of $P\in\ppp_\CC$
such that
$\operatorname{Ad}_{\mathrm{T}}(P)=a(T) P$ for all $T \in\Ttt$.
One can show that the spaces $\ppp_a$ are one-dimensional complex
vector spaces. The appearing characters $a \in\Phi\subset X^*(\Ttt)$
are called roots of $\Kkk$.
If $a \in\Phi$ is a root, then also $-a \in\Phi$.
Note that the character $-a$ as a map on $\Ttt$
is given by $(-a)(T)=(a(T))^{-1}$.

One can divide the vector space $\Vv$ in an upper half space and a
lower half space in
such a way
that there is no root on the boundary. A root in the upper half space is
then called a positive root. The set of vectors $v \in\Vv$ that satisfy
$\langle v, a \rangle\geq0$ for all positive roots $a$ is a so-called
positive Weyl chamber $\mathcal{C}_+$.
An element of the lattice $X^*(\Ttt)$ lying in
the positive Weyl chamber is called
a highest weight. The set of highest weights will be denoted by $\Ww_+$.
There is a one-to-one correspondence between the irreducible
representations and the highest
weight vectors.
\begin{theo}\label{theo-highest-weights}
Any irreducible (unitary)
representation of $\Kkk$ induces (by restriction)
a representation of $\Ttt$,
which when decomposed into irreducible representations of $\Ttt$ contains
exactly one highest weight $a\in\Ww_+$. For any highest weight vector
$a\in\Ww_+$,
there is exactly one irreducible representation of $\Kkk$
containing $a$.
\end{theo}

Let $\pi_a\dvtx  \Kkk\to\mathrm{U}(d(a))$ for $a\in\Ww_+$ be the corresponding
irreducible unitary representation of dimension $d(a)$.
By Schur orthogonality and the Peter--Weyl theorem
the matrix coefficients $\pi_a(K)_{k,l} $, where
$1\leq k,l \leq d(a)$, of these representations,
considered as functions on $\Kkk$, form an orthogonal basis for
$L^2(\Kkk)$.
The $L^2$ norm of such a matrix coefficient is $d(a)^{-1/2}$.
Therefore, the orthogonal projection of $f$ onto the space spanned by the
matrix coefficients of the irreducible representation $\pi_a$
is given by
\begin{eqnarray*}
\sum_{k,l=1}^{d(a)}
\int_\Kkk{d}\tilde{K} ( f(\tilde{K}) \overline{\pi
_a(\tilde
{K})_{k,l}} )
\pi_a(K)_{k,l}
& = &
\sum_{k,l=1}^{d(a)} \int_\Kkk{d}\tilde{K}
( f(\tilde{K}) \pi_a(\tilde{K}^{-1})_{l,k} ) \pi_a(K)_{k,l}
\\
& = &
d(a) \Tr(\Ff f(a) \pi_a(K) ) ,
\end{eqnarray*}
where
\[
\Ff f(a)
= \int_\Kkk{d}K\, f(\tilde{K}) \pi_a(K^{-1}) .
\]
Hence Schur orthogonality and the Peter--Weyl theorem imply the following.
\begin{coro}\label{coro-fourier}
Let $f \in L^2(\Kkk)$,
then one obtains with convergence in $L^2(\Kkk)$
\[
f(K)
=
\sum_{a \in\Ww_+} d(a) \Tr(\Ff f(a) \pi_a(K) ) .
\]
\end{coro}

As shown in \cite{Su}, one can characterize the smooth functions on
$\Kkk$
by their Fourier series.
%
\begin{theo}\label{theo-Su}
A function $f$ on $\Kkk$ is smooth if and only if its Fourier coefficients
are rapidly decreasing, which means that
\[
\forall h>0 \dvtx
\lim_{\|a\| \to\infty}
\|a\|^h \|\Ff f(a)\|
=0.
\]
Here $\|\Ff f(a)\|$ denotes the Hilbert--Schmidt norm.
If this is fulfilled, then the Fourier series converges absolutely
in the supremum norm on $\Kkk$.
\end{theo}

Note that the definition of $\Ff f(a)$ to be rapidly decreasing is
independent of the chosen norm on $\Ww_+\subset\Vv$.

Now let us consider the compact group $\Kkk\times\TM^{L}$ with the maximal
torus $\Ttt\times\TM^{L}$ and its Lie algebra
$\mathfrak{t}\times\RR^L$.
The characters of this torus also factorize by
$X^*(\Ttt\times\TM^{L})=X^*(\Ttt)\times\ZZ^L$.
As $\{\one\}\times\TM^{L}$ lies in the center, the direct product of
the scalar product on $\mathfrak{k}$ and the canonical scalar product
on $\RR^L$ give a scalar product on $\mathfrak{k}\times\RR^L$ that
is invariant
under the adjoint representation of the group $\Kkk\times\TM^L$.
Therefore, the induced scalar product on the vector space
$\Vv\times\RR^L$ spanned by the characters also factorizes.

As the adjoint representation of $\{\one\} \times\TM^L$ is trivial,
the roots of $\Kkk\times\TM^L$ consist of elements
$(a,0)$ where $a$ is a root of $\Kkk$.
Therefore, the positive roots of $\Kkk\times\TM^L$ are simply the
positive roots
of $\Kkk$ and, as the scalar product on $\Vv\times\RR^L$ factorizes,
the positive Weyl chamber for $\Kkk\times\TM^L$ is given by
$\mathcal{C}_+
\times\RR^L$.
Hence, the highest weight vectors are given by $\Ww_+ \times\ZZ^L$.

Now for $a\in\Ww_+$ the mapping
$(K,\theta) \mapsto\pi_a(K) e^{\imath j \cdot\theta}$ is an irreducible
representation of $\Kkk\times\TM^L$ which contains the
highest weight vector $(a,j)$ and by Theorem~\ref{theo-highest-weights},
it is the unique one containing this weight.
Thus, we have shown the following.
\begin{theo}
\label{theo-K-with-torus}
The highest weight vectors of $\Kkk\times\TM^L$ are given by
$\Ww_+ \times\ZZ^L$, where $\Ww_+$ are the highest weight vectors
of $\Kkk$. The irreducible representation parameterized by
$(a,j) \in\Ww_+ \times\ZZ^L$ is given by
\[
\pi_{(a,j)} (K,\theta)
=
\pi_a(K) e^{\imath j \cdot\theta} .
\]
Hence, the Fourier series of $F$ is given by
\[
F (K,\theta)
=
\sum_{a\in\Ww_+} \sum_{j\in\ZZ^{L}} d(a) \Tr(\Ff F(a,j)
\pi_{a}(K) ) e^{\imath j \cdot\theta}
\]
with convergence in $L^2(\Kkk\times\TM^L)$, where
\[
\Ff F(a,j)
=
\int_{\Kkk} {d}K \int_{\TM^{L}} {d}\theta\, F(K,\theta)
\pi_{a}(K^{-1})
e^{-\imath j\cdot\theta} .
\]
\end{theo}
\vspace*{-8pt}

\section{Divergence of vector fields}\label{appC}

Let $\Hhh\subset\Kkk$ be some compact subgroups
of the unitary group $\mathrm{U}(L)$ and
let $\Mm=\Kkk/ \Hhh$ be the homogeneous quotient and $\pi\dvtx \Kkk\to
\Mm$.
On the Lie algebra $\mathrm{u}(L)$ and hence on the Lie algebra $\mathfrak
{k}$ of
$\Kkk$,
the Killing form $(u,v)=\Tr(u^* v)$
defines a bi-invariant metric.
At each point $K\in\Kkk$, the Lie algebra $\mathfrak{h}$ of $\Hhh$
form the
vertical vectors, that is, the kernel of the
differential of $\pi$.
Hence, the tangent space at $\pi(K)$ can be identified with the
horizontal vectors, $\mathfrak{h}^\perp$, the orthogonal
complement of $\mathfrak{h}$ in $\mathfrak{k}$. This identification
depends on
the choice of $K$. Two horizontal lifts of some tangent vector on $\Mm$
to two different
preimages differ by a conjugation and therefore have the same length
due to the invariance of
the metric.
Thus, there is a unique metric on $\Mm$ such that the projection $\pi\dvtx
\Kkk\to\Mm$
is a Riemannian submersion. This metric is invariant under the action
of $\Kkk$.

Let $S_i$ be some orthonormal basis for $\mathfrak{h}^\perp$, then
the push forward, $\pi_*(S_i)$ forms an orthonormal basis at $\pi(K)$.
(This basis vectors may differ for two different preimages.)
Let $X$ be some smooth vector field on $\Mm$ and denote the horizontal
lift to $\Kkk$
by $\hat{X}$ which then is also smooth.
As $\pi$ is a Riemannian submersion, the covariant derivative
of $X$ with respect to $\pi_*(S_i)$ is given by $\pi_*(\nabla_{S_i}
\hat X)$.
Let $(B_j)$ denote some orthonormal basis of $\mathfrak{k}$ and identify
$B_j$ with the left invariant vector field.
Furthermore, we identify any vector field $Y$ with a function
$Y\dvtx  \Kkk\to\mathfrak{k}$ such that the vector at $K$ is given by
the path $K\exp(tY(K))$.
With $\nabla_S \hat{X}$, we denote the covariant derivative of the
vector field
$\hat{X}$ and with $\delta_S \hat{X}$ the derivative
of the function w.r.t. to the left-invariant vector field $S$.
Then one has
\[
\nabla_{S} \hat{X}
=
\sum_{j} \nabla_{S} \Tr(B_j^*\hat{X}) B_j
=
\sum_j \biggl[
\Tr(B_j^*\hat{X})\frac{1}{2} [S, B_j] + \delta_{S} \hat{X}
\biggr].
\]
If $g$ denotes the metric on $\Mm$, then the divergence of $X$ at $\pi
(K)$ is given by
\[
\operatorname{div}(X)\circ\pi
=
\sum_i g(\pi_*(S_i), \nabla_{\pi_*S_i} X) \circ\pi
=
\sum_i \Tr(S_i^* \nabla_{S_i} \hat{X}),
\]
where we used that $S_i$ is horizontal so that
$g(\pi_*(S_i),\pi_*(Y))=\Tr(S_i^* Y)$ for all $Y$.
Using the identity above and the fact that
$S_i^*=-S_i$ which implies $\Tr(S_i^* [S_i, B_j])=0$,
the expression reduces to
%
%
\begin{equation}
\label{eq-div}
\operatorname{div}(X)\circ\pi
=
\sum_i \delta_{S_i} \Tr(S_i^* \hat{X}).
\end{equation}
As $\Tr(S_i^* Y)=0$ for any vertical vector $Y\in\mathfrak{h}$,
the lifted vector field $\hat{X}$ does not need to be horizontal for the
last equation to hold.
\end{appendix}

\section*{Acknowledgments}
We thank A. Bendikov for many helpful discussions and hints to the
literature at an early stage of this work, and the referee for a number
of helpful suggestions.

%

%
\printaddresses

\end{document}